# Unikraft: Fast, Specialized Unikernels the Easy Way

Simon Kuenzer
NEC Laboratories Europe GmbH

Vlad-Andrei Bădoiu*
University Politehnica of Bucharest

Hugo Lefeuvre*
The University of Manchester

Sharan Santhanam*
NEC Laboratories Europe GmbH

Alexander Jung*
Lancaster University

Gaulthier Gain*
University of Liège

Cyril Soldani*
University of Liège

Costin Lupu
University Politehnica of Bucharest

Ștefan Teodorescu
University Politehnica of Bucharest

Costi Răducanu
University Politehnica of Bucharest

Cristian Banu
University Politehnica of Bucharest

Laurent Mathy
University of Liège

Răzvan Deaconescu
University Politehnica of Bucharest

Costin Raiciu
University Politehnica of Bucharest

Felipe Huici
NEC Laboratories Europe GmbH

## Abstract

Unikernels are famous for providing excellent performance in terms of boot times, throughput and memory consumption, to name a few metrics. However, they are infamous for making it hard and extremely time consuming to extract such performance, and for needing significant engineering effort in order to port applications to them. We introduce Unikraft, a novel micro-library OS that (1) fully modularizes OS primitives so that it is easy to customize the unikernel and include only relevant components and (2) exposes a set of composable, performance-oriented APIs in order to make it easy for developers to obtain high performance.

Our evaluation using off-the-shelf applications such as nginx, SQLite, and Redis shows that running them on Unikraft results in a 1.7x-2.7x performance improvement compared to Linux guests. In addition, Unikraft images for these apps are around 1MB, require less than 10MB of RAM to run, and boot in around 1ms on top of the VMM time (total boot time 3ms-40ms). Unikraft is a Linux Foundation open source project and can be found at www.unikraft.org.

## 1 Introduction

Specialization is arguably the most effective way to achieve outstanding performance, whether it is for achieving high throughput in network-bound applications [38, 50, 52], making language runtime environments more efficient [20, 23, 47, 65], or providing efficient container environments [62, 76], to give some examples. Even in the hardware domain, and especially with the demise of Moore's law, manufacturers are increasingly leaning towards hardware specialization to achieve ever better performance; the machine learning field is a primary exponent of this [30, 32, 34].

In the virtualization domain, unikernels are the golden standard for specialization, showing impressive results in terms of throughput, memory consumption, and boot times, among others [36, 40, 45, 47, 48]. Some of those benefits come from having a single memory address space, thus eliminating costly syscall overheads, but many of those are the result of being able to hook the application at the right level of abstraction to extract best performance: for example, a web server aiming to service millions of requests per second can access a low-level, batch-based network API rather than the standard but slow socket API. Such an approach has been taken in several unikernel projects but often in an ad hoc, build-and-discard manner [38, 48, 52]. In all, despite clear benefits, unikernels suffer from two major drawbacks:

- They require significant expert work to build and to extract high performance; such work has to, for the most part, be redone for each target application.
- They are often non-POSIX compliant, requiring porting of applications and language environments.

We argue that these drawbacks are not fundamental, and propose a unikernel architecture built specifically to address them. Existing unikernel projects, even those based on library architectures, tend to consist of small but monolithic kernels that have complex, intertwined and sometimes opaque APIs for their components. This means that developers not only have to often port applications to such systems, but that optimizing their performance requires digging into the code and the specifics of the (uni)kernel in order to understand how to best obtain performance gains.

*These are main authors that contributed equally to the paper.

Further, such systems typically rely on size-based specialization: removing all unnecessary components to achieve minimal images. While this strategy already offers significant benefits, we argue that unikernels based on library architectures should *ease* access to true specialization, allowing users to choose the best system component for a given application, environmental constraints, and key performance indicators.

In this paper we propose Unikraft, a novel *micro-library* operating system targeted at painlessly and seamlessly generating specialized, high performance unikernels. To do so, Unikraft relies on two key principles:

- The kernel should be fully modular in order to allow for the unikernel to be fully and easily customizable. In Unikraft, OS primitives such as memory allocators, schedulers, network stacks and early boot code are stand-alone micro-libraries.
- The kernel should provide performance-minded, well-defined APIs that can be easily selected and composed in order to meet an application's performance needs. In Unikraft, such APIs are micro-libraries themselves, meaning that they can be easily added to or removed from a build, and that their functionality can be extended by providing additional such micro-libraries.

In brief, the key conceptual innovation of Unikraft is defining a small set of APIs for core OS components that makes it easy to replace-out a component when it is not needed, and to pick-and-choose from multiple implementations of the same component when performance dictates. The APIs have been built with performance (e.g., by supporting batching by design) and minimality in mind (no unneeded features).

To support a wide range of applications, we port the musl libc library, and provide a syscall shim layer micro-library. As a result, running an application on Unikraft can be as simple as building it with its native build system, and linking the resulting object files back into Unikraft. In addition, Unikraft supports a number of already-ported applications (e.g., SQLite, nginx, Redis), programming languages and runtime environments such as C/C++, Go, Python, Ruby, Web Assembly and Lua, and a number of different hypervisors/VMMs (QEMU/KVM, Xen, Firecracker [4], and Solo5 [78] as of this writing).

Our evaluation using such applications on Unikraft results in a 1.7x-2.7x performance improvement compared to Linux guests. In addition, Unikraft images for these apps are around 1MB, require less than 10MB of RAM to run, and boot in around 1ms on top of the VMM time (total boot time 2ms-40ms). Unikraft is a Linux Foundation open source project and the sources can be found at www.unikraft.org.

## 2 Design Principles and Solution Space

Before deriving what the key design principles for Unikraft are, it is worth analyzing the features and (heavyweight) mechanisms of traditional OSes that are unnecessary or ill-suited to single application use cases:

- Protection-domain switches between the application and the kernel might be redundant in a virtualization context because isolation is ensured by the hypervisor, and result in measurable performance degradation.
- Multiple address spaces may be useless in a single application domain, but removing such support in standard OSes requires a massive reimplementation effort.
- For RPC-style server applications, threading is not needed, with a single, run-to-completion event loop sufficing for high performance. This would remove the need for a scheduler within the VM and its associated overheads, as well as the mismatch between the guest and hypervisor schedulers [19].
- For performance-oriented UDP-based apps, much of the OS networking stack is useless: the app could simply use the driver API, much like DPDK-style applications already do. There is currently no way to easily remove just the network stack but not the entire network sub-system from standard OSes.
- Direct access to NVMe storage from apps removes the need for file descriptors, a VFS layer and a filesystem, but removing such support from existing OSes, built around layers of the storage API, is very difficult.
- Memory allocators have a large impact on application performance, and general purpose allocators have been shown to be suboptimal for many apps [66]. It would therefore be ideal if each app could choose its own allocator; this is however very difficult to do in today's operating systems because the allocators that kernels use are baked in.

This admittedly non-exhaustive list of application-specific optimizations implies that for each core functionality that a standard OS provides, there exists at least one or a few applications that do not need it. Removing such functionality would reduce code size and resource usage but would often require an important re-engineering effort.

The problem we want to solve is to enable developers to create a specialized OS for every single application to ensure the best performance possible, while at the same time bounding OS-related development effort and enabling easy porting of existing applications. This analysis points to a number of key design decisions:

- **Single address space**: Target single application scenarios, with possibly different applications talking to each other through networked communications.
- **Fully modular system**: All components, including operating system primitives, drivers, platform code and libraries should be easy to add and remove as needed; even APIs should be modular.
- **Single protection level**: There should be no user-/kernel-space separation to avoid costly processor mode

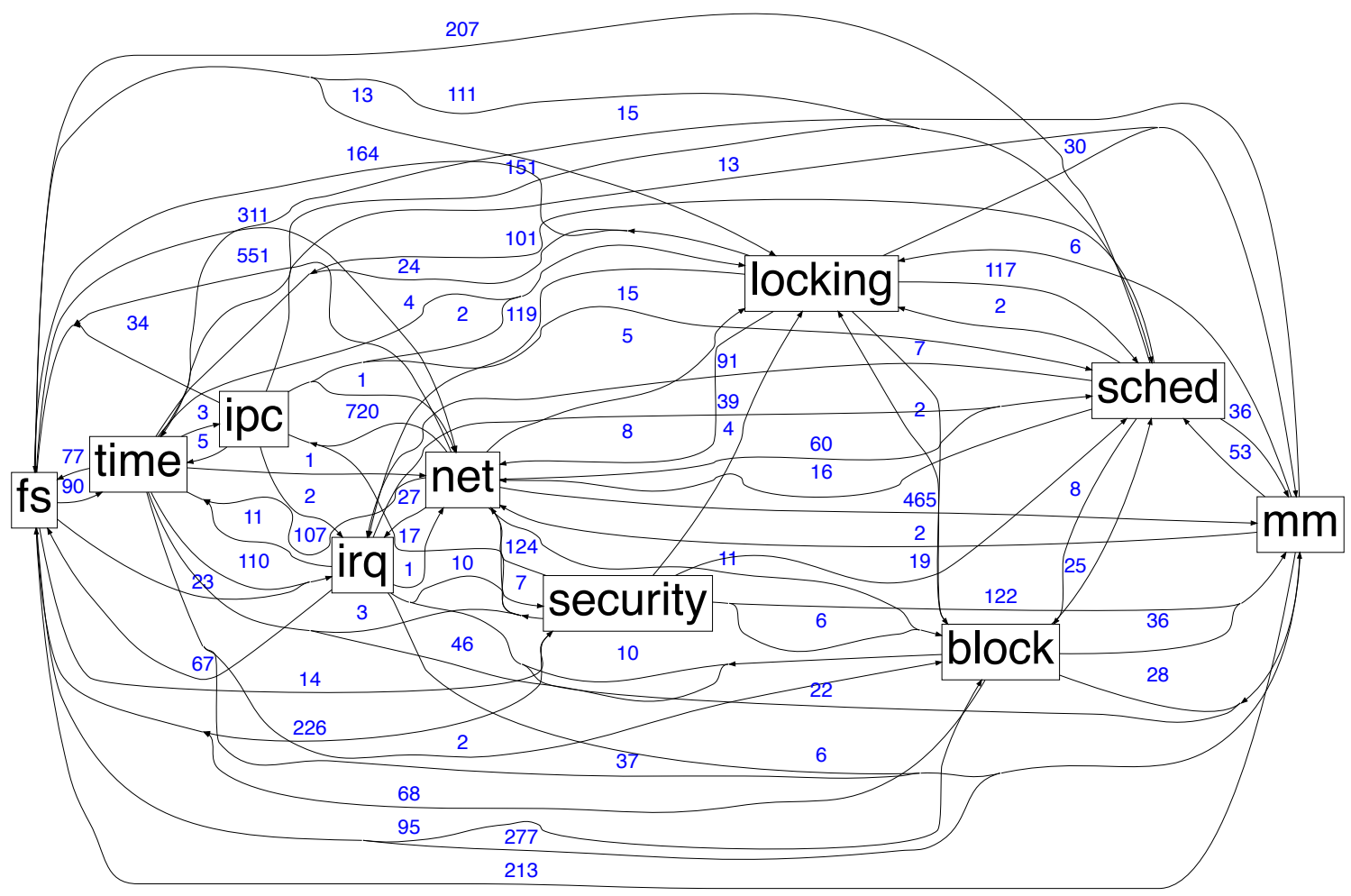

**Figure 1.** Linux kernel components have strong inter-dependencies, making it difficult to remove or replace them.

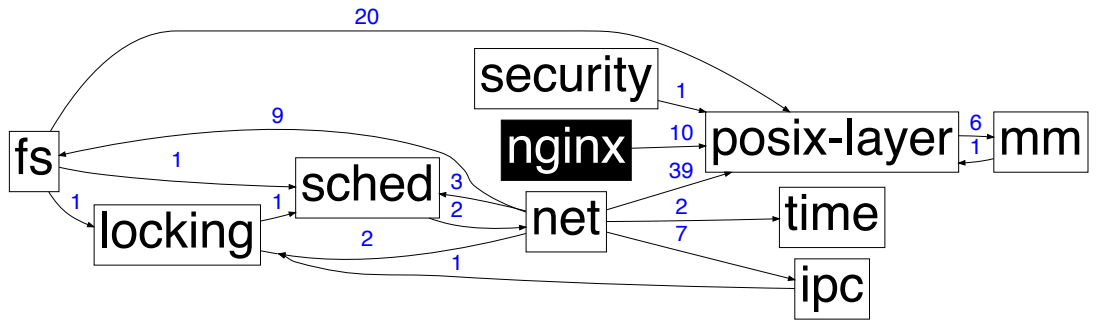

**Figure 2.** Nginx Unikraft dependency graph

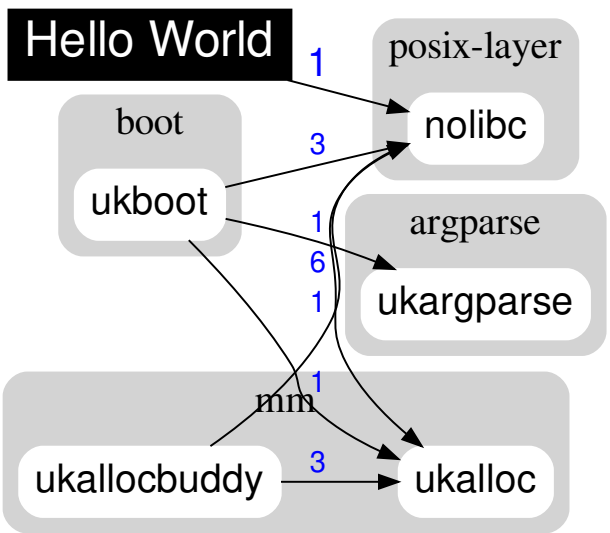

**Figure 3.** Helloworld Unikraft dependency graph

switches. This does not preclude compartmentalization (e.g., of micro-libraries), which can be achieved at reasonable cost [69].

- **Static linking**: Enable compiler features, e.g., Dead Code Elimination (DCE) and Link-Time Optimization (LTO), to automatically get rid of unneeded code.
- **POSIX support**: In order to support existing or legacy applications and programming languages while still allowing for specialization under that API.
- **Platform abstraction**: Seamless generation of images for a range of different hypervisors/VMMs.

Given these, the question is how to implement such a system: by minimizing an existing general-purpose operating system, by starting from an existing unikernel project, or from scratch.

Existing work has taken three directions in tackling this problem. The first direction takes existing OSes and adds or removes functionality. Key examples add support for a single address space and remove protection domain crossings: OSv [37] and Rump [36] adopt parts of the BSD kernel and re-engineer it to work in a unikernel context; Lupine Linux [40] relies on a minimal, specialized configuration of the Linux kernel with Kernel Mode Linux (KML) patches. These approaches make application porting easy because they provide binary compatibility or POSIX compatibility, but the resulting kernel is monolithic.

Existing monolithic OSes do have APIs for each component, but most APIs are quite rich as they have evolved organically, and component separation is often blurred to achieve performance (e.g., `sendfile` short circuits the networking and storage stacks). The Linux kernel, for instance, historically featured highly inter-dependent subsystems [8].

To better quantify this API complexity, we analyzed dependencies between the main components of the Linux kernel. As a rough approximation, we used the subdirectories in the kernel source tree to identify (broad) components. We used `cscope` to extract all function calls from the sources of all kernel components, and then for each call checked to see if the function is defined in the same component or a different one; in the latter case, we recorded a dependency. We plot the dependency graph in Figure 1: the annotations on the edges show the number of dependencies between nodes. This dense graph makes it obvious that removing or replacing any single component in the Linux kernel requires understanding and fixing all the dependencies of other components, a daunting task.

While full modularization is difficult, modularizing certain parts of a monolithic kernel has been done succesfully by Rump. There, the NetBSD kernel was split into base layers (which must be used by all kernels), functions provided by the host (scheduling, memory allocation,etc) and so-called *factions* that can be run on their own (e.g. network or filesystem support). Rump goes some way towards achieving our goals, however there are still many dependencies left which require that all kernels have the base and hypercall layers.

The second direction is to bypass the operating system (OS) altogether, mostly for I/O performance, while leaving the original stack in place – wasting resources in the process. Even here, porting effort is required as apps must be coded against the new network (DPDK, netmap [64] or Linux's `io_uring` [11] subsystem) or storage (SPDK) API.

The third direction is to add the required OS functionality from scratch for each target application, possibly by reusing code from existing operating systems. This is the approach taken by ClickOS [51] to support Click modular routers, MirageOS [46] to support OCaml applications, and MiniCache [39] to implement a web cache, to name a few. The resulting images are very lean, have great performance and have small boot times; the big problem is that the porting effort is huge, and that it has to be mostly repeated for every single application or language.

In sum, starting from an existing project is suboptimal since none of the projects in the three directions mentioned were designed to support the key principles we have outlined. We opt for a clean-slate API design approach, though we do reuse components from existing works where relevant.

## 3 Unikraft Architecture and APIs

In contrast to classical OS work, which can be roughly split between monolithic kernels (with great performance) versus micro-kernels that provide great isolation between OS components (at the expense of performance), our work embraces both the monolithic design (no protection between components) and the modularity that micro-kernels advocated.

We use modularity to enable specialization, splitting OS functionality into fine-grained components that only communicate across well-defined API boundaries. Our key observation is that we can obtain performance via careful API design and static linking, rather than short-circuiting API boundaries for performance. To achieve the overarching principle of modularity, Unikraft consists of two main components:

- **Micro-libraries**: Micro-libraries are software components which implement one of the core Unikraft APIs; we differentiate them from libraries in that they have minimal dependencies and can be arbitrarily small, e.g., a scheduler. All micro-libraries that implement the same API are interchangeable. One such API contains multiple memory allocators that all implement the ukalloc interface. In addition, Unikraft supports libraries that can provide functionality from external library projects (OpenSSL, musl, Protobuf [31], etc.), applications (SQLite, Redis, etc.), or even platforms (e.g., Solo5, Firecracker, Raspberry Pi 3).
- **Build system**: This provides a Kconfig-based menu for users to select which micro-libraries to use in an application build, for them to select which platform(s) and CPU architectures to target, and even configure individual micro-libraries if desired. The build system then compiles all of the micro-libraries, links them, and produces one binary per selected platform.

Figure 4 shows Unikraft's architecture. All components are micro-libraries that have their own Makefile and Kconfig configuration files, and so can be added to the unikernel build independently of each other[1]. APIs are also micro-libraries that can be easily enabled or disabled via a Kconfig menu; unikernels can thus compose which APIs to choose to best cater to an application's needs (e.g., an RCP-style application might turn off the `uksched` API in order to implement a high performance, run-to-completion event loop).

[1]Unless, of course, a micro-library has a dependency on another, in which case the build system also builds the dependency.

Unikraft's architecture also includes components that add POSIX support, making it relatively easy to support existing applications (more on this in §4). Unikraft can improve the performance of applications in two ways:

1. **Unmodified applications**, by eliminating syscall overheads, reducing image size and memory consumption, and by choosing efficient memory allocators.
2. **Specialization**, by adapting applications to take advantage of lower level APIs wherever performance is critical (e.g., a database application seeking high disk I/O throughput).

As a proof of Unikraft's modularity, a minimal Hello World configuration yields an image of 200KB in size on KVM and 40KB on Xen, requiring only platform boostrapping code and `nolibc`, a Unikraft-specific `libc` replacement that only provides a basic minimal set of functionality such as `memcpy` and string processing. All the libraries used in this image are shown in Figure 3; in contrast, the entire Linux kernel in Figure 1 is needed for a Hello World app in Linux.

Most applications do require more functionality (see nginx image in Figure 2). Note how (1) this image does not include a block subsystem since it only uses RamFS, and (2) how all components are smaller and have fewer dependencies than their Linux counterparts. These examples showcase Unikraft's ability to easily add and remove components, including core OS ones, allowing developers to create efficient, specialized images for their apps.

The ability to easily swap components in and out, and to plug applications in at different levels presents application developers with a wide range of optimization possibilities. To begin with, unmodified applications (e.g. Hello World and nginx) can use the posix-compatibility layer with musl (① in Figure 4) or nolibc, transparently getting low boot times, lower memory consumption and improved throughput because of the lack of syscall overheads, as Unikraft syscalls are effectively function calls.

Likewise, the application developer can easily select an appropriate memory allocator (⑥) to obtain maximum performance, or to use multiple different ones within the same unikernel (e.g., a simple, fast memory allocator for the boot code, and a standard one for the application itself).

Developers interested in fast boot times could further optimize the unikernel by providing their own boot code (⑤) to comply with the `ukboot` API; in §6 we show experiments with two boot code micro-libraries, one with static memory pages and one with dynamic ones, showing the trade-off between boot time and memory allocation flexibility.

For network-bound applications, the developers can use the standard socket interface (②) or the lower level, higher performance `uknetdev` API (⑦) in order to significantly improve throughput; we will discuss this API in greater detail below, and will evaluate it in §6. Similarly, disk-bound applications such as databases can follow a standard path through the `vfscore` micro-library (③), or optimize throughput by coding against the `ukblock` API (⑧). Schedulers are also pluggable (④), and each CPU core can run a different scheduler (even if multi-core support is still work in progress).

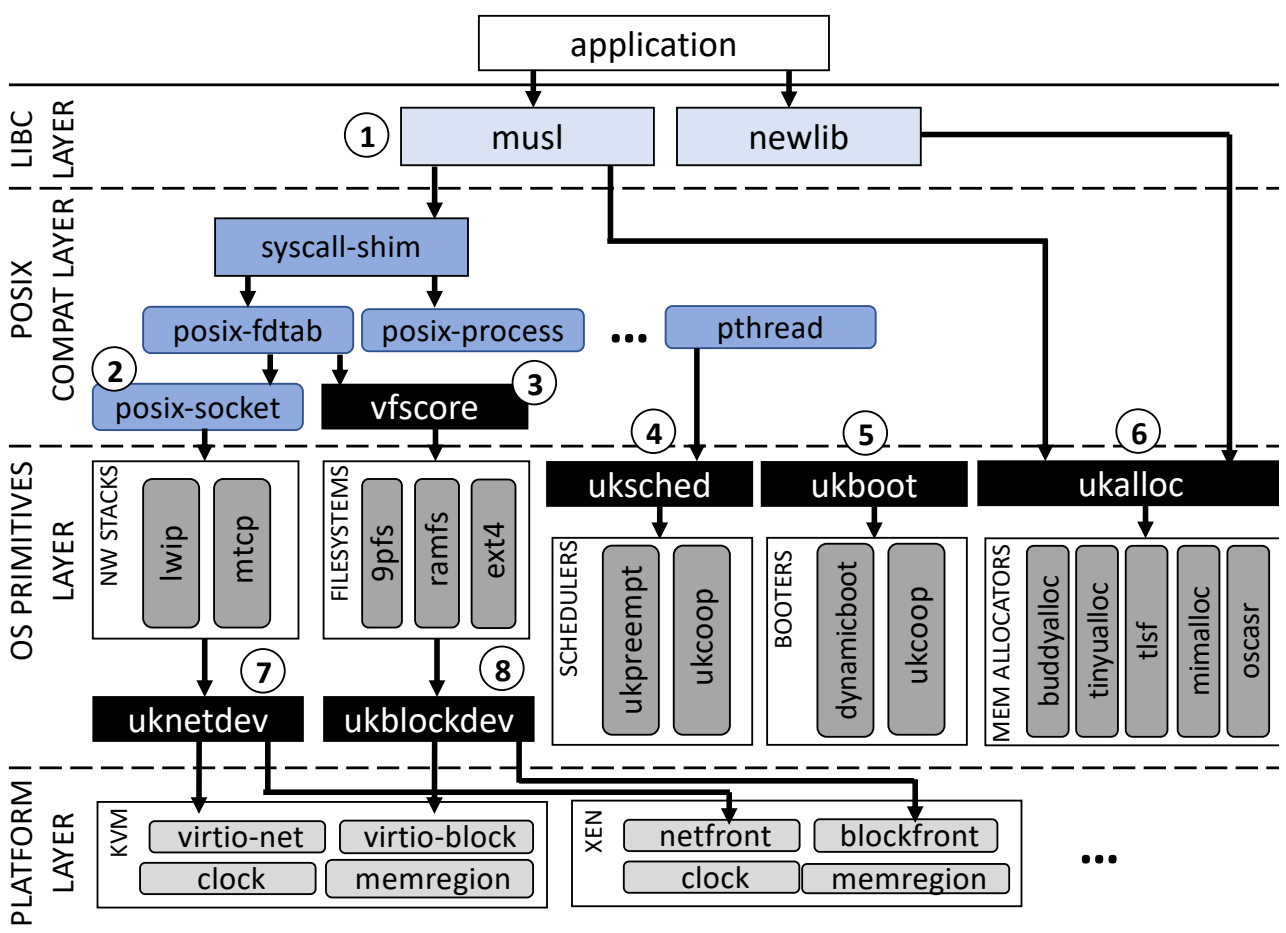

**Figure 4.** The Unikraft architecture (APIs in black boxes) enables specialization by allowing apps to plug into APIs at different levels and to choose from multiple API implementations.

We will visit and evaluate several of these scenarios later on in the paper, but first we give a more in-depth look into Unikraft's APIs by focusing on a subset of them.

### 3.1 uknetdev API

Unikraft's networking sub-system decouples the device driver side (e.g., virtio-net, netfront) from the network stack or low-level networking application (application for short).

Regarding the former, easily swapping network stacks is something that is not common in commodity OSes; instead, drivers are usually implemented for a particular network stack. The aim of this API is to decouple these two components in order to allow drivers to be reused across platforms.

For the latter, a networking application or network stack should be able to run unmodified on a different platform with different drivers. Because we are addressing a wide range of use cases, the API should not restrict any of them nor become a potential performance bottleneck for high performance workloads. We derived part of the design from Intel DPDK's `rte_netdev` API. However, because its focus is on high performance rather than efficient resource usage, we designed an API that allows applications to operate Unikraft drivers in polling, interrupt-driven, or mixed mode.

In addition, `uknetdev` leaves memory management to the application, all the while supporting high performance features like multiple queues, zero-copy I/O, and packet batching. We let the application fully operate and initialize the driver; drivers do not run any initialization routine on their own. Instead, we provide API interfaces for applications to provide necessary information (e.g., supported number of queues and offloading features) so that the application code can specialize by picking the best set of driver properties and features. Drivers register their callbacks (e.g, send and receive) to a `uk_netdev` structure which the application then uses to call the driver routines.

In order to develop application-independent network drivers while using the application's or network stack's memory management we introduce a network packet buffer wrapper structure called `uk_netbuf`. This struct contains meta information needed by the driver to send or receive data in the packet buffer; the idea is that applications use this structure for packet buffer allocations, but the layout is under the control of the application. Since neither the driver nor the API manage allocations, performance critical workloads can make use of pre-allocated network buffers pools, while memory efficient applications can reduce memory the footprint by allocating buffers from the standard heap. The send and receive calls of the API look as follows:

```
int uk_netdev_tx_burst(struct uk_netdev *dev,
                       uint16_t queue_id,
                       struct uk_netbuf **pkt,
                       __u16 *cnt);

int uk_netdev_rx_burst(struct uk_netdev *dev,
                       uint16_t queue_id,
                       struct uk_netbuf **pkt,
                       __u16 *cnt);
```

The user hands over arrays of `uk_netbufs` and specifies their length. On transmit, the driver enqueues as many netbufs as possible from the given array (size is given with `cnt`). The function returns flags that indicate if there is still room on the queue to send more packets or if the queue is full. The `cnt` parameter is also used as an output parameter to indicate how many packets were actually placed on the send queue. The receive function works in a similar manner.

As default, a driver operates a queue in polling mode, but the API has one more interface to enable interrupt mode for a specific queue. In this mode, whenever the function indicates that there is no more work to be done (no more packets received or the send queue is full), the interrupt line of the queue is enabled. During driver configuration the application can register an interrupt handler per queue which is called as soon as a packet is received or space becomes available on the transmit queue. Afterwards, the interrupt line is inactive until the transmit or receive function activates it again according to the queue state, for instance when all packets were received by the application. How and where packets are received or transmitted is entirely up to the application's implementation. For instance, the interrupt callback could be used to unblock a receiving or sending thread, but could also be included into the eventloop implementations. As soon as an interrupt arrives the application knows that the receive or send function has to be called. This implementation avoids interrupt storms and enables automatic transition to polling mode under heavy load situations.

### 3.2 ukalloc API

Unikraft's memory allocation subsystem is composed of three layers: (1) a POSIX compliant external API, (2) an internal allocation API called ukalloc, and (3) one or more backend allocator implementations. The external interface is motivated by backward compatibility to facilitate the porting of existing applications to Unikraft. In the case of the C language, the external API is exposed by a modified standard library which can be nolibc (a minimal, Unikraft-specific libc implementation), newlib or musl. The external allocation interface acts as a compatibility wrapper for the Unikraft-specific internal allocation interface, which in turn redirects allocation requests to the appropriate allocator backend (each allocator has its own, separate memory region). The internal allocation interface therefore serves as a multiplexing facility that enables the presence of multiple memory allocation backends within the same unikernel.

Unikraft's allocation interface exposes uk_ prefixed versions of the POSIX interface: uk_malloc(), uk_calloc(), etc. In contrast to POSIX, these functions require the caller to specify which allocation backend should be used to satisfy the request. uk_malloc() is defined as:

```
static inline void *
uk_malloc (struct uk_alloc *a, size_t size);
```

The struct uk_alloc * argument represents the allocation backend. This structure contains function pointers that refer to the allocator's implementation of the POSIX allocation interface: malloc(), calloc(), posix_memalign(), etc. Note that uk_malloc(), like most of the internal allocation interface, is designed as an inline method in order to avoid any additional function call overhead in the allocation path.

Allocators must specify an initialization function which is called by ukboot at an early stage of the boot process. Initialization functions are passed a void * base pointer to the first usable byte of the heap, along with a size_t len argument which specifies the size of the heap. They must fully initialize the allocator and register the allocator with the ukalloc interface. The allocator is considered ready to satisfy memory allocations as soon as the initialization function returns. The boot process sets the association between memory allocators and memory sources.

Unikraft supports five allocation backends: a buddy system, the Two-Level Segregated Fits [53] (TLSF) real-time memory allocator, tinyalloc [67], Mimalloc [42] (version 1.6.1) and the Oscar [12] secure memory allocator. A special case are garbage-collection (GC) memory allocators that require a thread to perform GC. We can implement these with two allocators, one for the early boot time that initializes the GC thread, and then the main GC allocator, which takes over as soon as its thread is started; we use this solution for Mimalloc because it has a pthread dependency.

### 3.3 uksched and uklock APIs

Unlike many OSes, scheduling in Unikraft is available but optional; this enables building lightweight single-threaded unikernels or run-to-completion unikernels, avoiding the jitter caused by a scheduler within the guest. Example use cases are to provide support functions as virtual machines (as in driver domains), or Virtual Network Functions (VNFs).

Similar to ukalloc, uksched abstracts actual scheduler interfaces. The platform library provides only basic mechanisms like context switching and timers so that scheduling logic and algorithms are implemented with an actual scheduler library (Unikraft supports co-operative and pre-emptive schedulers as of this writing).

Additionally, Unikraft supports instantiating multiple schedulers, for example one per available virtual CPU or for a subset of available CPUs. For VNFs for example, one may select no scheduling or cooperative scheduling on virtual cores that run the data plane processing because of performance and delay reasons, but select a common preemptive scheduler for the control plane.

The uklock library provides synchronization primitives such as mutexes and semaphores. In order to keep the code of other libraries portable, uklock selects a target implementation depending on how the unikernel is configured. The two dimensions are threading and multi-core support. In the simplest case (no threading and single core), some of the primitives can be completely compiled out since there is no need for mutual exclusion mechanisms. If multi-core were enabled (we do not yet support this), some primitives would use spin-locks and RCUs, and so in this case they would be compiled in.

## 4 Application Support and Porting

Arguably, an OS is only as good as the applications it can actually run; this has been a thorn on unikernels' side since their inception, since they often require manual porting of applications. More recent work has looked into using binary compatibility, where unmodified binaries are taken and syscalls translated, at run-time, into a unikernel's underlying functionality [37, 54]. This approach has the advantage of requiring no porting work, but the translation comes with important performance penalties.

| Platform | Routine call | #Cycles | nsecs |
|---|---|---|---|
| *Linux/KVM* | System call | 222.0 | 61.67 |
| | System call (no mitigations) | 154.0 | 42.78 |
| *Unikraft/KVM* | System call | 84.0 | 23.33 |
| *Both* | Function call | 4.0 | 1.11 |

**Table 1.** Cost of binary compatibility/syscalls with and without security mitigations.

To quantify these, Table 1 shows the results of microbenchmarks ran on an Intel i7 9700K 3.6 GHz CPU and Linux 5.11 that compare the cost of `no-op` system and function calls in Unikraft and Linux (with run-time syscall translation for Unikraft). System calls with run-time translation in Unikraft are 2-3x faster than in Linux (depending on whether KPTI and other mitigations are enabled). However, system calls with run-time translation have a tenfold performance cost compared to function calls, making binary compatibility as done in OSv, Rump and HermiTux [36, 37, 54] expensive.

For virtual machines running a single application, syscalls are likely not worth their costs, since isolation is also offered by the hypervisor. In this context, unikernels can get important performance benefits by removing the user/kernel separation and its associated costs. The indirection used by binary compatibility reduces unikernel benefits significantly.

To avoid these penalties but still minimize porting effort, we take a different approach: we rely on the target application's native build system, and use the statically-compiled object files to link them into Unikraft's final linking step. For this to work, we ported the musl C standard library, since it is largely glibc-compatible but more resource efficient, and newlib, since it is commonly used to build unikernels.

To support musl, which depends on Linux syscalls, we created a micro-library called *syscall shim*: each library that implements a system call handler registers it, via a macro, with this micro-library. The shim layer then generates a system call interface at *libc*-level. In this way, we can link to system call implementations directly when compiling application source files natively with Unikraft, with the result that syscalls are transformed into inexpensive function calls.

Table 2 shows results when trying this approach on a number of different applications and libraries when building against `musl` and newlib: this approach is not effective with newlib [2] ("std" column), but it *is* with `musl`: most libraries build fully automatically. For those that do not, the reason has to do with the use of `glibc`-specific symbols (note that this is not the case with newlib, where many `glibc` functions are not implemented at all). To address this, we build a `glibc` compatibility layer based on a series of `musl` patches [56] and 20 other functions that we implement by hand (mostly 64-bit versions of file operations such as `pread` or `pwrite`).

With this in place, as shown in the table ("compat layer" column), this layer allows for almost all libraries and applications to compile and link. For `musl` that is good news: as long as the syscalls needed for the applications to work are implemented, then the image will run successfully (for newlib the stubs would have to be implemented).

## 4.1 Application Compatibility

How much syscall support does Unikraft have? As of this writing, we have implementations for 146 syscalls; according to related work [54, 74], in the region of 100-150 syscalls are enough to run a rich set of mainstream applications, frameworks and languages; we confirm this in Table 3, which lists software currently supported by Unikraft.

| | `musl` | | | `newlib` | | | glue |
|---|---|---|---|---|---|---|---|
| | Size (MB) | std | compat. layer | Size (MB) | std | compat. layer | code LoC |
| lib-axtls | 0.364 | ✗ | ✓ | 0.436 | ✗ | ✓ | 0 |
| lib-bzip2 | 0.324 | ✗ | ✓ | 0.388 | ✗ | ✓ | 0 |
| lib-c-ares | 0.328 | ✗ | ✓ | 0.424 | ✗ | ✓ | 0 |
| lib-duktape | 0.756 | ✓ | ✓ | 0.856 | ✗ | ✓ | 7 |
| lib-farmhash | 0.256 | ✓ | ✓ | 0.340 | ✓ | ✓ | 0 |
| lib-fft2d | 0.364 | ✓ | ✓ | 0.440 | ✗ | ✓ | 0 |
| lib-helloworld | 0.248 | ✓ | ✓ | 0.332 | ✓ | ✓ | 0 |
| lib-httpreply | 0.252 | ✓ | ✓ | 0.372 | ✗ | ✓ | 0 |
| lib-libucontext | 0.248 | ✓ | ✓ | 0.332 | ✗ | ✓ | 0 |
| lib-libunwind | 0.248 | ✓ | ✓ | 0.328 | ✓ | ✓ | 0 |
| lib-lighttpd | 0.676 | ✗ | ✓ | 0.788 | ✗ | ✓ | 6 |
| lib-memcached | 0.536 | ✗ | ✓ | 0.660 | ✗ | ✓ | 6 |
| lib-micropython | 0.648 | ✓ | ✓ | 0.708 | ✗ | ✓ | 7 |
| lib-nginx | 0.704 | ✗ | ✓ | 0.792 | ✗ | ✓ | 5 |
| lib-open62541 | 0.252 | ✓ | ✓ | 0.336 | ✓ | ✓ | 13 |
| lib-openssl | 2.9 | ✗ | ✓ | 3.0 | ✗ | ✓ | 0 |
| lib-pcre | 0.356 | ✓ | ✓ | 0.432 | ✗ | ✓ | 0 |
| lib-python3 | 3.1 | ✗ | ✓ | 3.2 | ✗ | ✓ | 26 |
| lib-redis-client | 0.660 | ✗ | ✓ | 0.764 | ✗ | ✓ | 29 |
| lib-redis-server | 1.3 | ✗ | ✓ | 1.4 | ✗ | ✓ | 32 |
| lib-ruby | 5.6 | ✗ | ✓ | 5.7 | ✗ | ✓ | 37 |
| lib-sqlite | 1.4 | ✗ | ✓ | 1.4 | ✗ | ✓ | 5 |
| lib-zlib | 0.368 | ✗ | ✓ | 0.432 | ✗ | ✓ | 0 |
| lib-zydis | 0.688 | ✓ | ✓ | 0.756 | ✗ | ✓ | 0 |

**Table 2.** Automated porting using externally-built archives linked against Unikraft using musl and newlib.

Beyond this, we conduct a short analysis of how much more work it might take to support additional applications. We use the Debian popularity contest data [14] to select a set of the 30 most popular server applications (*e.g.,* apache, mongodb, postgres, avahi, bind9). To derive an accurate set of syscalls these applications require to actually run, and to extend the static analysis done in previous work [61] with dynamic analysis, we created a small framework consisting of various configurations (*e.g.,* different port numbers for web servers, background mode, *etc.*) and unit tests (*e.g.,* SQL queries for database servers, DNS queries for DNS servers, *etc.*). These configurations and unit tests are then given as input to the analyzer which monitors the application's behavior by relying on the *strace* utility. Once the dynamic analysis is done, the results are compared and added to the ones from the static analysis.

We plot the results against the syscalls currently supported by our system in the heatmap on Figure 5 (the entire analysis and heatmap generation is fully automated by a set of tools we developed). Each square represents an individual syscall, numbered from 0 (`read`) to 313 (`finit_module`). Lightly colored squares are required by none of the applications (0 on the scale) or few of them (20% of them); black squares (e.g., square 1, `write`) are required by all. A number on a square means that a syscall is supported by Unikraft, and an empty square is a syscall not supported yet.

As can be seen from the map, more than half the syscalls are not even needed in order to support popular applications,

| | |
|---|---|
| **Applications** | NGINX, SQLite, Redis, memcached, Click modular router, lighttpd (ongoing). |
| **Frameworks** | Intel DPDK, TensorFlow Lite, PyTorch. |
| **Compiled Languages** | C/C++, Go, Web Assembly (WAMR), Lua, Java/OpenJDK (ongoing), Rust (ongoing) |
| **Interpreted Languages** | Python, Micropython, Ruby, JavaScript/v8 (ongoing). |

**Table 3.** Applications, frameworks and languages currently supported by Unikraft.

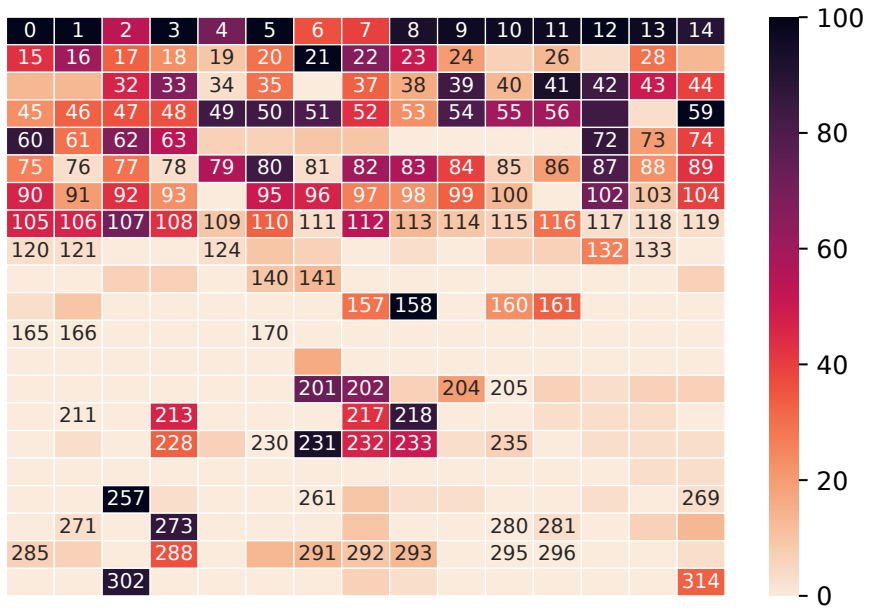

**Figure 5.** Syscalls required by 30 server apps vs syscalls supported by Unikraft.

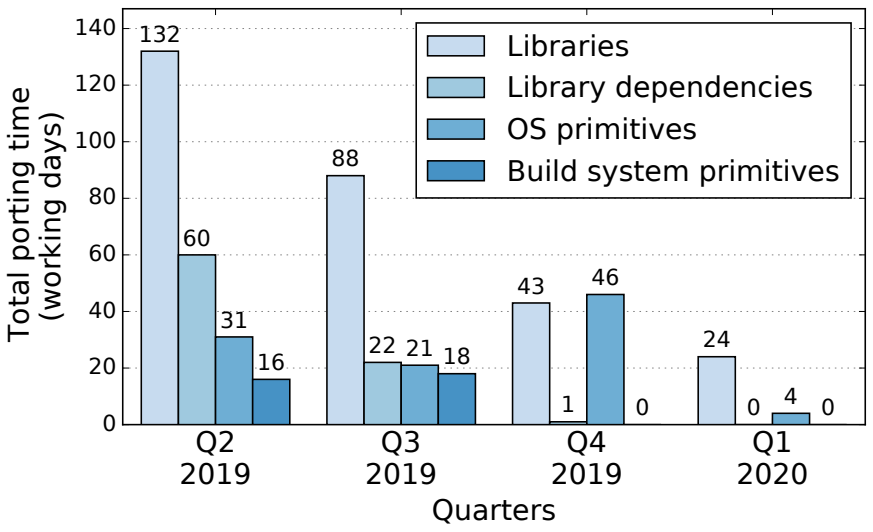

**Figure 6.** Devel survey of total effort to port a library, including dependencies, missing OS and build system primitives.

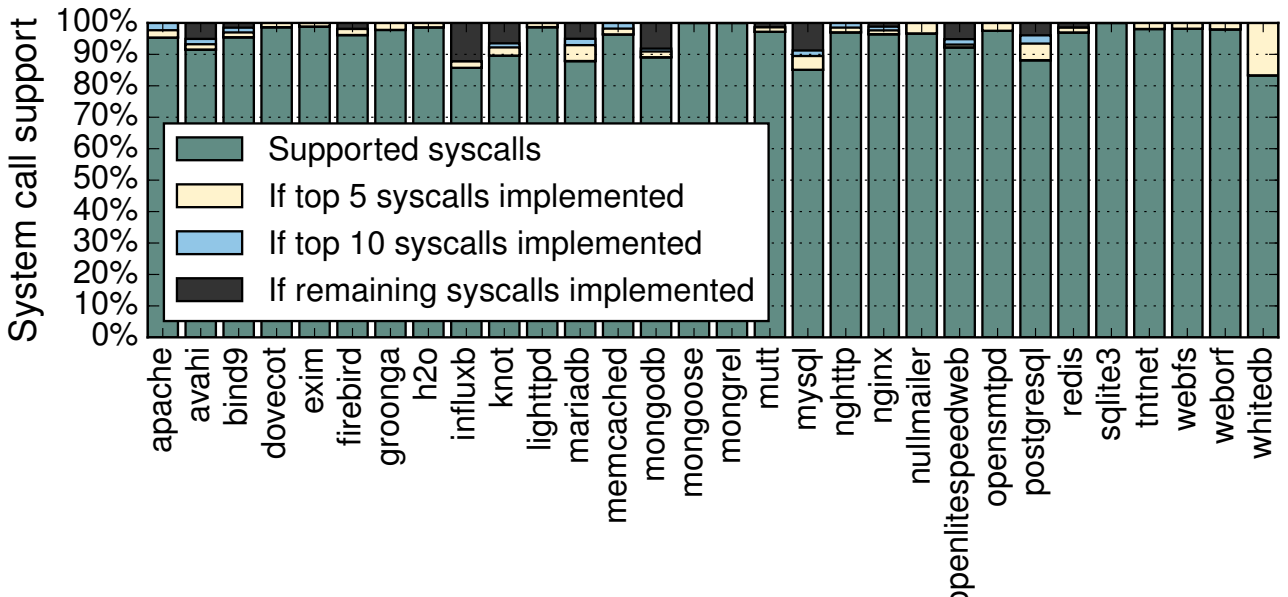

**Figure 7.** Syscall support for top 30 server apps [14]. All apps are close to being supported, and several already work even if some syscalls are stubbed (SQLite, nginx).

and most of the needed syscalls we already support. Of those that are not supported (in the order of about 60):

- several can be quickly stubbed in a unikernel context (e.g., `getcpu`, if using a single cpu);
- many are relatively trivial to implement since the necessary functionality is already supported by Unikraft (e.g., `semget`/`semopt`/`semctl`).
- and the rest are work in progress (e.g., `epoll`, `eventfd`).

To quantify this even further, Figure 7 plots, for each of the selected 30 applications, how much of their needed syscalls Unikraft supports so far (in green), how close to full support we would be if we implemented the next 5 most common syscalls across all 30 applications (yellow), the next 10 (light blue), all the way to full support. The first take-away is that all applications are close to having full support (the graph is mostly green). The second thing to note is that even applications that we do have running are not all green (e.g., SQLite, nginx): this is because many applications work even if certain syscalls are stubbed or return `ENOSYS` (which our shim layer automatically does if a syscall implementation is missing). We are currently in the process of conducting a survey of how many of these applications can actually run despite their bars not being completely green.

We estimate that a moderate level of additional engineering work to support these missing syscalls would result in even wider support for applications. Finally, for cases where the source code is not available, Unikraft also supports binary compatibility and binary rewriting as done in HermiTux [54].

## 4.2 Manual Porting

When the automated approach does not work and performance is paramount so binary compatibility is not an option, we rely on manual porting. Note, however, that because Unikraft's micro-libraries provide a common code base for building specialized stacks, such manual porting is significantly less time consuming than that of past unikernels projects (including our own work) that take in the order of months to be put together; this is clear from the few lines of glue code we needed to add when porting a wide range of libraries and applications (see Table 2, last column).

In fact, anecdotal accounts throughout the lifetime of Unikraft point to the fact that, as the common code base has matured, porting additional functionality has gotten increasingly easier. Admittedly, quantifying the actual man hours spent porting a library is a difficult exercise (e.g., because commit timestamps may hide the fact that, during the course of porting a library, significant time was spent porting one of its dependencies). Nevertheless, we have conducted a survey of all developers in the project's open source community (around 70) who have ported a library or application, and present the results here. In particular, we asked developers to roughly calculate the time it took to port an actual library or application, the time it took to port library dependencies (e.g., memcached requires libevent), and the time it took to implement missing OS primitives (e.g., the `poll()` function) or add functionality to Unikraft's build system. We use git commit history to track when a port was started.

To show the evolution of the porting effort as the project matured, we plot the results of the survey in a time-line starting in March 2019 and ending in May 2020 in Figure 6; for ease of presentation, we further group the results in quarters. The figure confirms the anecdotal evidence that, as time progressed, the amount of time developers had to spend porting dependencies or implementing missing OS primitives has significantly decreased.

Finally, Unikraft's support for a wide range of languages and their environments (standard libraries, garbage collectors, *etc.*) means that a number of projects based on these (*e.g.,* Intel's DNNL/C++, Django/Python, Ruby on Rails/Ruby, etc.) should work out of the box.

## 5 Base Evaluation

The main goal of Unikraft is to help developers quickly and easily create resource-efficient, high-performance unikernels. In this section we evaluate to what extent Unikraft achieves this goal *transparently*, i.e., without having to modify applications (essentially scenarios 1-3 in the architecture diagram in Figure 4); then, in Section 6 we will evaluate how (slight) modifications to comply with Unikraft's APIs can result in even higher performance.

Throughout our evaluation, we use KVM as the virtualization platform. Unikraft also supports Xen and bare-metal targets (e.g., Raspberry Pi and Xilinx Ultra96-V2), but we leave their performance evaluation to future work. We run all experiments on an inexpensive (roughly €800) Shuttle SH370R6 computer with an Intel i7 9700K 3.6 GHz (4.9 Ghz with Turbo Boost, 8 cores) and 32GB of RAM. For the DPDK experiment we use two of these connected via a direct cable and a pair of Intel X520-T2 cards with the 82599EB chipset.

Further, we disabled Hyper-Threading and isolated 4 CPU cores for the host using kernel boot parameters (isolcpus=4-7 noht); from the remaining 4 CPU cores we pinned one to the VM, another one to the VMM (e.g., qemu-system), and another one to the client tool (e.g., wrk or redis-benchmark), and set the governor to performance. Finally, for the Xen boot experiments we use Xen version 4.0.

All experiments were conducted by pinning a CPU core to the VM, another one to the VMM (e.g., `qemu-system`), and another one to the client tool (e.g., `wrk` or `redis-benchmark`); by disabling Hyper-threading; and by setting the governor to performance.

### 5.1 Resource Efficiency: Smaller is Better

The main advantage of unikernels over traditional OSes is their low resource consumption. This is reflected in binary image size when on disk, and boot-time and memory footprint at runtime. We evaluated these for a number of representative apps in Unikraft, in comparison with leading unikernels, and Linux. Our results are shown in figs. 8 to 11.

In order to quantify image sizes in Unikraft, we generate a number of images for all combinations of DCO and LTE, and for a helloworld VM and three other applications: nginx, Redis and SQLite. The results in Figure 8 show that Unikraft images are all under 2MBs for all of these applications. We further compare these results with other unikernels and Linux in Figure 9. As shown, Unikraft images are smaller than all other unikernel projects and comparable to Linux userspace binaries (note that the Linux sizes are just for the application; they do not include the size of glibc nor the kernel). This is a consequence of Unikraft's modular approach, drastically reducing the amount of code to be compiled and linked (e.g., for helloworld, no scheduler and no memory allocator are needed).

Small image sizes are not only useful for minimizing disk storage, but also for quick boot times. LightVM [48] has shown that it is possible to boot a no-op unikernel in around 2ms, with a heavily optimized Xen toolstack. In our evaluation, we use standard virtualization toolstacks instead, and wish to understand how quickly Unikraft VMs can boot. When running experiments, we measure both the time taken by the VMM (e.g. Firecracker, QEMU, Solo5) and the boot time of the actual unikernel/VM, measured from when the first guest instruction is run until `main()` is invoked.

The results are shown in Figure 10, showing how long a helloworld unikernel needs to boot with different VMMs. Unikraft's boot time on QEMU and Solo5 (guest only, without VMM overheads) ranges from tens (no NIC) to hundreds of microseconds (one NIC). On Firecracker, boot times are slightly longer but do not exceed 1ms. These results compare positively to previous work: MirageOS (1-2ms on Solo5), OSv (4-5ms on Firecracker with a read-only filesystem), Rump (14-15ms on Solo5), Hermitux (30-32ms on uHyve), Lupine (70ms on Firecracker, 18ms without KML), and Alpine Linux (around 330ms on Firecracker). This illustrates Unikraft's ability to only keep and initialize what is needed.

Overall, the total VM boot time is dominated by the VMM, with Solo5 and Firecracker being the fastest (3ms), QEMU microVM at around 10ms and QEMU the slowest at around 40ms (we elaborate on guest boot times in figs. 14 and 21). These results show that Unikraft can be readily used in scenarios where just-in-time instantiation of VMs is needed.

Finally, previous work [48] stressed the importance of not only fast instantiation but also VM density. In order to understand how many unikernel VMs we could pack on a single server when RAM is a bottleneck, we ran experiments to measure the minimum amount of memory required to boot various applications as unikernels, finding that 2-6MBs of memory suffice for Unikraft guests (Figure 11).

### 5.2 Filesystem Performance

Unikraft has a VFS layer which apps can link against for file I/O. Typically, Unikraft guests include a RAM filesystem when they do not require access to persistent storage. To support persistent storage, apps can use the 9pfs [77] protocol to access such storage on the host or in the network. Our 9pfs implementation relies on virtio-9p as transport for KVM, implementing the standard VFS operations. Enabling the 9pfs device adds 0.3ms to the boot time of Unikraft VMs on KVM, and 2.7ms on Xen.

We measured file-access latency for both read and write; the 9pfs filesystem resides in the host, is 1GB in size and contains random data. Our Unikraft test application reads chunks of sizes 4K, 8K, 16K and 32K, measuring the latency for each read. To benchmark against a Linux VM, we used dd to transfer the same data to `/dev/null`. We estimate the latency from the dd output by dividing the total time by the number of blocks. The results in Figure 20 show that Unikraft achieves lower read latency and lower write latency with different block sizes and are considerably better than ones from the Linux VM.

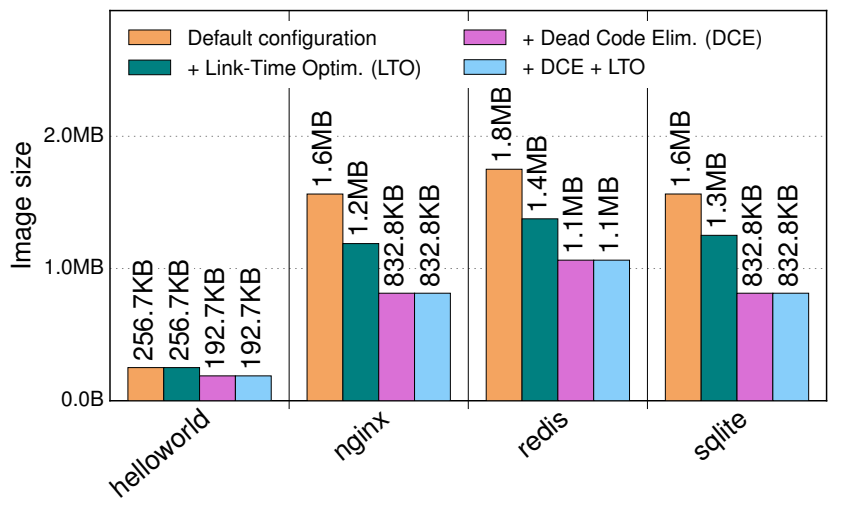

Figure 8. Image sizes of Unikraft applications with and without LTO and DCE.

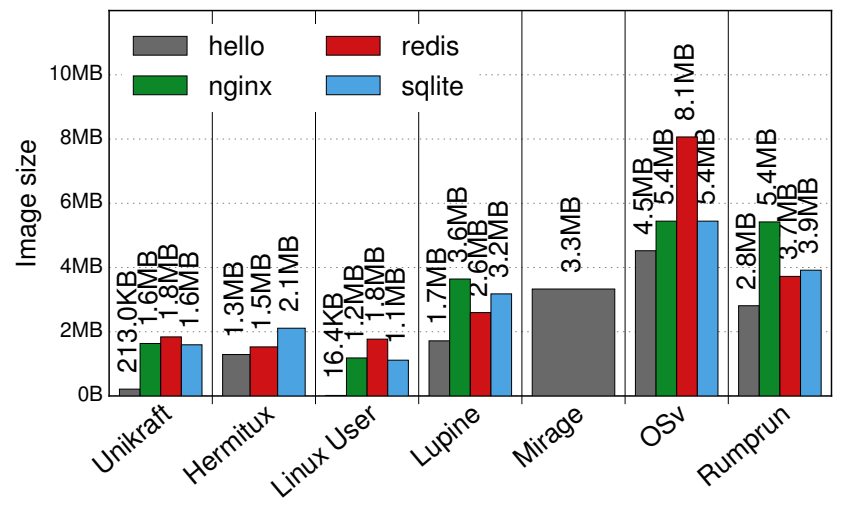

Figure 9. Image sizes for Unikraft and other OSes, stripped, w/o LTO and DCE.

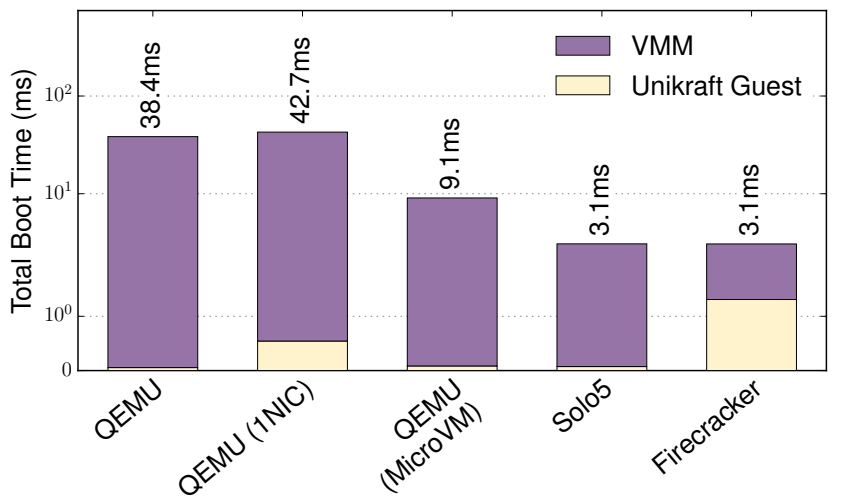

Figure 10. Boot time for Unikraft images with different virtual machine monitors.

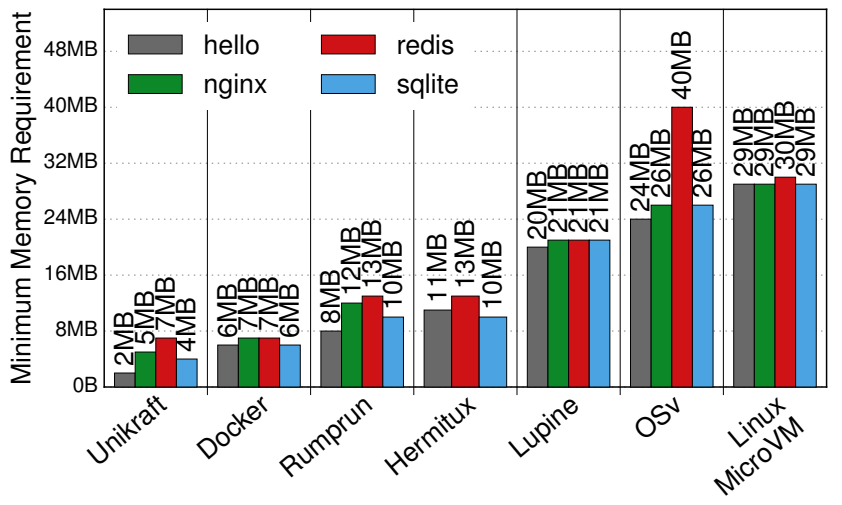

Figure 11. Minimum memory needed to run different applications using different OSes, including Unikraft.

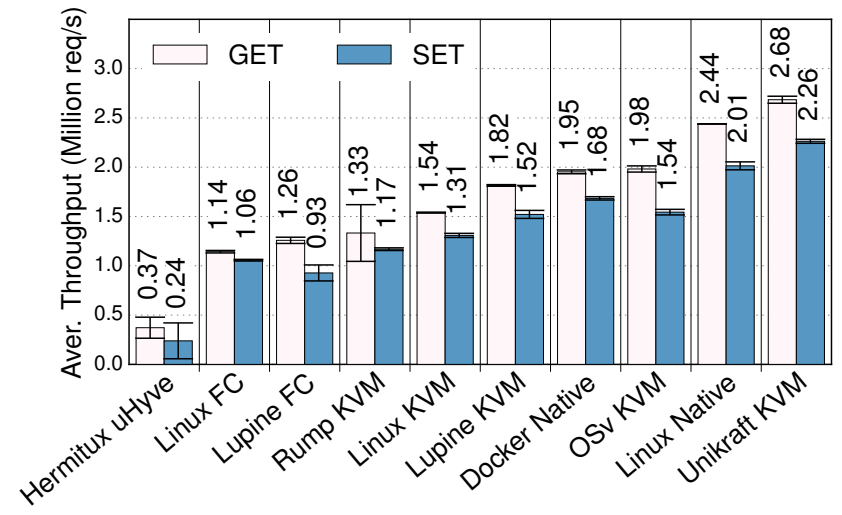

Figure 12. Redis perf (30 conns, 100k reqs, pipelining 16) with QEMU/KVM and Firecracker (FC).

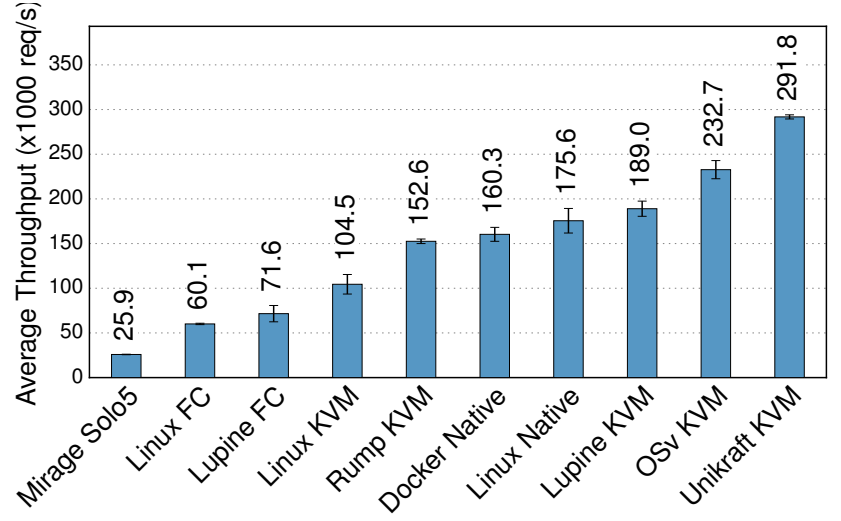

Figure 13. NGINX (and Mirage HTTP-reply) performance with `wrk` (1 minute, 14 threads, 30 conns, static 612B page).

## 5.3 Application Throughput

For this part of the evaluation, we use manual ports of nginx and Redis (two representative server apps) and compare their performance to that of existing solutions, including a non-virtualized (native) Linux binary, a Docker container, a Linux microVM, and a representative set of unikernels. We conduct all measurements with the same application config and where possible the same application version (this is limited by application support, *e.g.,* Lupine, HermiTux, and Rump only support a specific version of Redis), on a single core. We did not optimize application or kernel configs for performance, however we took care of removing obvious performance bottlenecks for each system, *e.g.,* switching on memory pools in Unikraft's networking stack (based on lwIP [17]), or porting Lupine to QEMU/KVM in order to avoid Firecracker performance bottlenecks [4, 24]. Unikraft measurements use Mimalloc as the memory allocator.

The results are shown in Figures 12 and 13. For both apps, Unikraft is around 30%-80% faster than running the same app in a container, and 70%-170% faster than the same app running in a Linux VM. Surprisingly, Unikraft is also 10%-60% faster than Native Linux in both cases. We attribute these results to the cost of system calls (aggravated by the presence of KPTI — the gap between native Linux and Unikraft narrows to 0-50% without KPTI), and possibly the presence of Mimalloc as system-wide allocator in Unikraft; unfortunately it is not possible to use Mimalloc as kernel allocator in Linux without heavy code changes. Note that we did try to `LD_PRELOAD` Mimalloc into Redis, but the performance improvement was not significant. We expect that the improvement would be more notable if Mimalloc were present at compile time instead of relying on the preloading mechanism (making the compiler aware of the allocator allows it to perform compile/link time optimizations) but we could not perform this experiment since Mimalloc is not natively supported by the current Redis code base.

Compared to Lupine on QEMU/KVM, Unikraft is around 50% faster on both Redis and NGINX. These results may be due to overcutting in Lupine's official configuration, scheduling differences (we select Unikraft's cooperative scheduler since it fits well with Redis's single threaded approach), or remaining bloat in Lupine that could not be removed via configuration options. Compared to OSv, Unikraft is about 35% faster on Redis and 25% faster for nginx. Rump exhibits poorer performance: it has not been maintained for a while, effectively limiting the number of configurations we could apply. For instance, we could not set the file limits because the program that used to do it (`rumpctrl`) does not compile anymore. HermiTux [55] does not support nginx; for Redis, its performance is rather unstable. This is likely due to the absence of virtio support, as well as performance bottlenecks at the VMM level (HermiTux relies on uHyve, a custom VMM that, like Firecracker, does not match the performance of QEMU/KVM). Unlike Lupine, porting it to QEMU/KVM requires significant code changes [25].

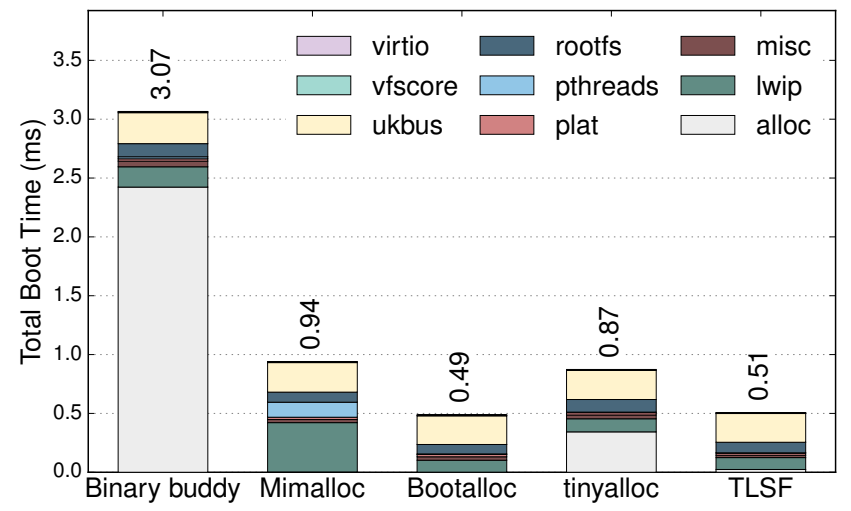

**Figure 14.** Unikraft Boot time for Nginx with different allocators.

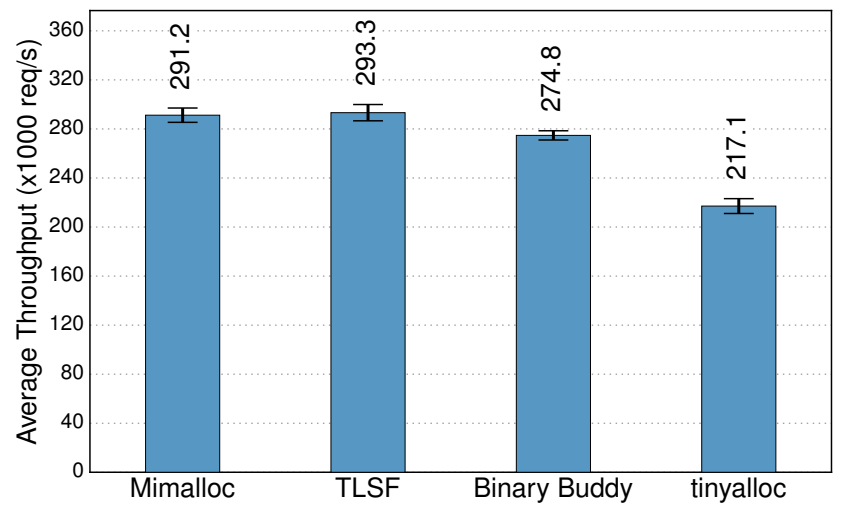

**Figure 15.** nginx throughput with different allocators.

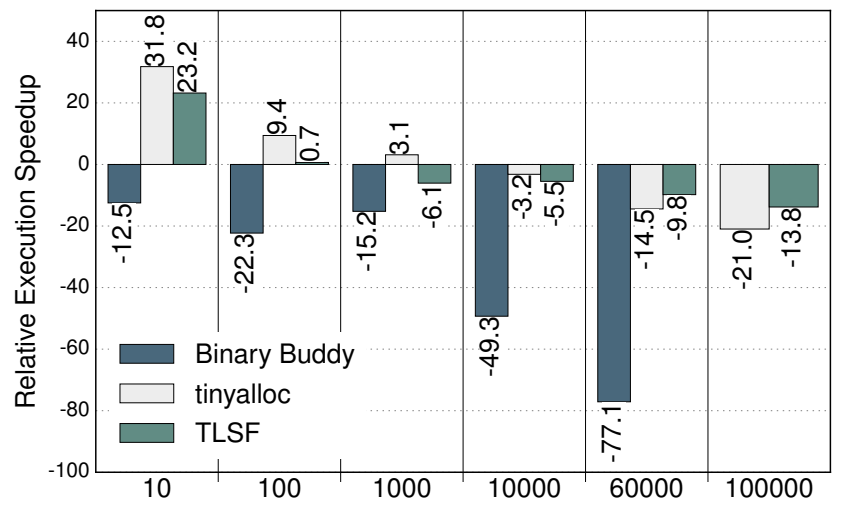

**Figure 16.** Execution speedup in SQLite Unikraft, relative to mimalloc [42].

### 5.4 Performance of Automatically Ported Apps

The apps tested so far have been ported manually. In this section, we provide an initial evaluation of automatic porting. We use SQLite and measure the time it takes to do 60K SQL insert queries, showing the results in Figure 17.

The manually ported versions of SQLite running on top of musl and newlib are denoted by the "native" bars in the graph, while the "external" bar shows the performance of SQLite with automatic porting; this means that we built SQLite using its own build system and linked the resulting static library with Unikraft, as discussed in §4. The results show that the automatically ported app is only 1.5% slower than the manually ported version, and even slightly faster than Linux baremetal (probably due to syscall overhead and the fact that we use TLSF as the memory allocator).

Overall, these results show that it is possible to have a mainstream application work in Unikraft with no porting effort (other than setting some compilation flags so that SQLite's build system generates a static library) and still reap the performance benefits of running it in a unikernel.

### 5.5 Memory Allocators

No single memory allocator is perfect for all purposes [66], making them a great target for specialization. To support this, the `ukalloc` API allows for multiple allocators to be present in a single image, and enables different micro-libraries to use different allocators (⑥ in the architecture from figure 4).

Unikraft has five memory allocators that comply with its API: (1) the buddy memory allocator from Mini-OS [41] [28]; (2) TLSF [53], a general purpose dynamic memory allocator specifically designed to meet real-time requirements; (3) mimalloc, a state-of-the-art, general-purpose allocator by Microsoft; (4) tinyalloc [67], a small and simple allocator; and (5) bootalloc, a simple region allocator for faster booting.

We built nginx with all the above allocators and measured the Unikraft guest boot time (Figure 14), as well as the sustained performance of the server (Figure 15). The difference in boot times for the different allocators is quite large: from 0.49ms (bootalloc) to 3.07ms (buddy), hinting that a just-in-time instantiation use-case of nginx should steer clear of the buddy allocator. At runtime, however, the buddy allocator performs similarly to tlsf and mimalloc, with tinyalloc taking a 30% performance hit.

Boot performance is similar for SQLite, with the buddy allocator being the worst and tinyalloc and tlsf among the best (results not shown for brevity). At runtime, though, the order depends on how many queries are run (see Figure 16): tinyalloc is fastest for less than 1000 queries by 3-30%, becoming suboptimal with more requests, as its memory compaction algorithms are slower; using mimalloc, instead, provides a 20% performance boost under high load.

Results for Redis (Figure 18), further confirm that no allocator is optimal for all workloads, and that the right choice of allocator for the workload and use-case can boost performance by 2.5x. In all, these results, and Unikraft's ability to concurrently support multiple allocators, leave room for future work into dynamically changing allocators based on current load in order to extract even better performance.

## 6 Specializing Applications

The previous section has shown that Unikraft core libraries are fast, and that by porting existing applications to Unikraft, we can outperform Linux.

Nevertheless, the power of Unikraft is in the ability to further customize images in ways that are difficult to achieve using existing solutions, be they monolithic OSes or other unikernel projects. The highly modular nature of Unikraft libraries makes it possible to easily replace core components such as memory allocators, page table support, schedulers, standard libraries and so on.

In this section we begin an initial exploration of the specialization opportunities Unikraft provides, with the main goal of achieving better performance. Other applications of specialization including better robustness, correctness or security, are the subject of future work. Throughout the section we will make reference to the architecture diagram (Figure 4) to point out which of the various specialization scenarios in it we are covering.

### 6.1 Specialized Boot Code (scenario 5)

As a first experiment, we try different implementations of guest paging support. By default, the Unikraft binary contains an already initialized page-table structure which is loaded in memory by the VMM; during boot Unikraft simply enables paging and updates the page-table base register to point to the right address. This is enough for most applications, and provides fast boot performance (30us boot for 1GB of RAM, see Figure 21). Unikraft also has dynamic page management support which can be enabled when apps need to alter their virtual address space explicitly (e.g. via `mmap`); when this is used the entire page-table is populated at boot time. Figure 21 shows that a guest with a 32MB dynamic page-table takes slightly longer to boot than one with a pre-initialized 1GB page-table, and that the boot time increases proportionally with the amount of memory.

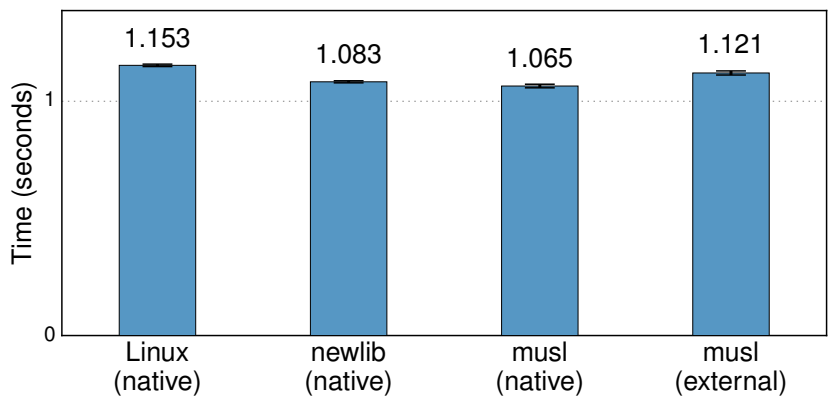

**Figure 17.** Time for 60k SQLite insertions for native Linux, newlib and musl on Unikraft and SQLite ported automatically to Unikraft (musl external).

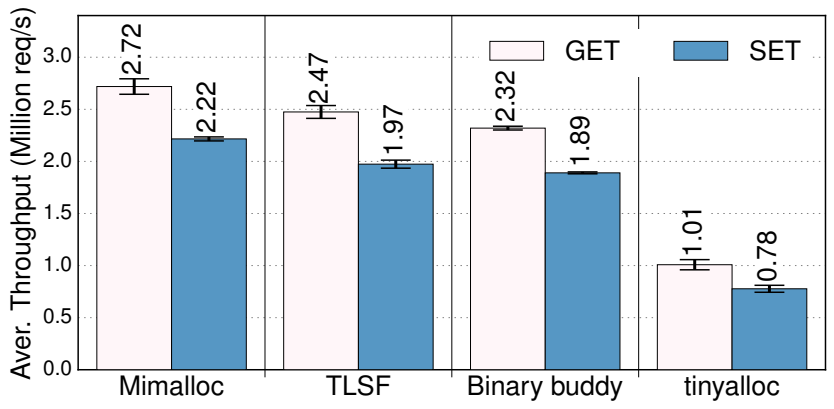

**Figure 18.** Redis throughput on Unikraft for different allocators (`redis-benchmark`, 30 conns, 100k requests, pipelining level of 16.)

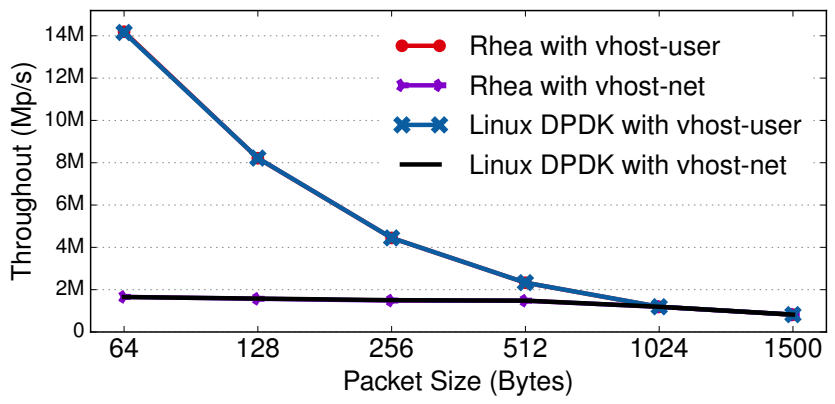

**Figure 19.** TX throughput comparison of Unikraft versus DPDK in a Linux VM for vhost-user aand vhost-net.

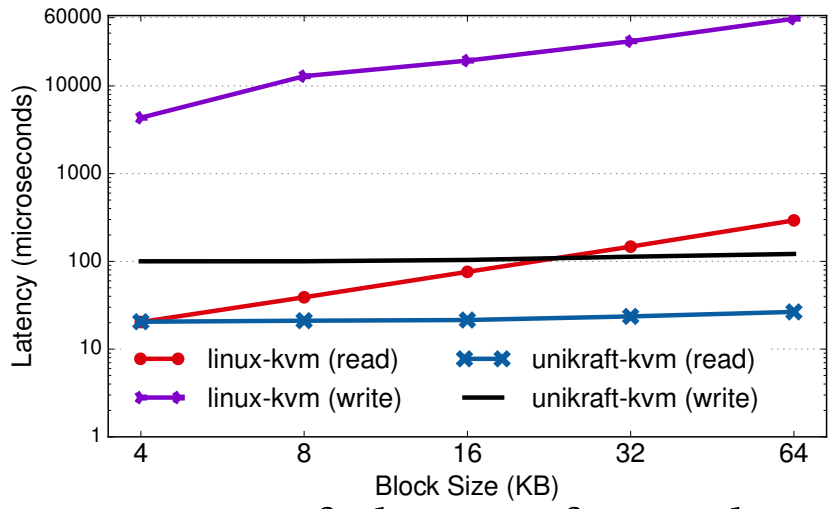

**Figure 20.** 9pfs latency for read and write operations, compared to Linux.

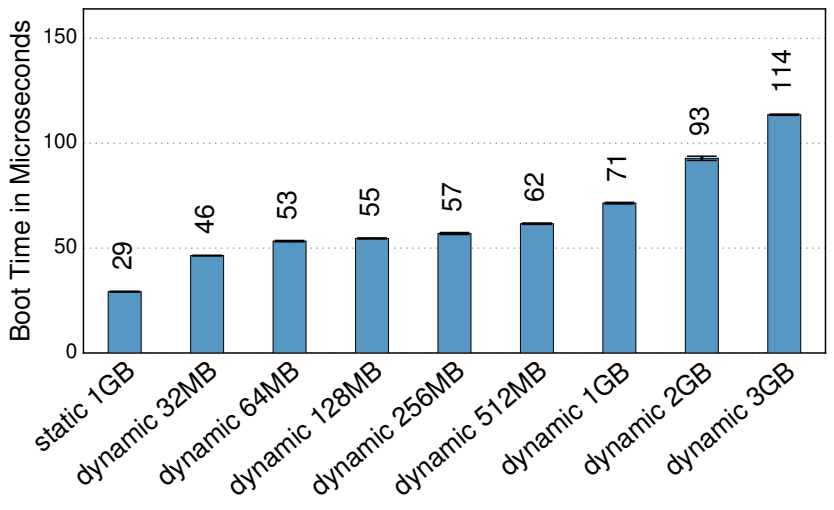

**Figure 21.** Unikraft boot times w/ static and dynamic page table initialization.

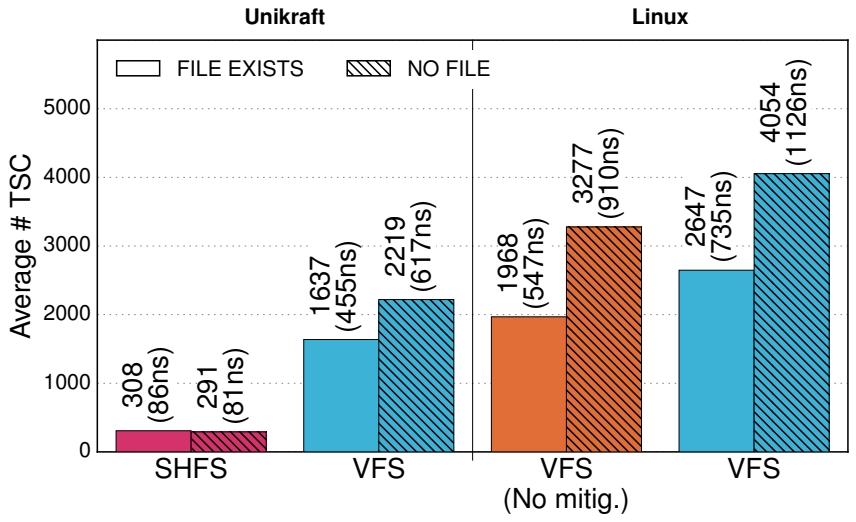

**Figure 22.** Perf. with a specialized filesystem and removing the VFS layer.

Finally, Unikraft makes it possible for the guest to run in protected (32 bit) mode, disabling guest paging altogether. This could help run legacy 32 bit applications, or reduce the cost of TLB misses in high-CPU contention scenarios.

## 6.2 Networking Performance (scenario 7)

The `uknetdev` API is one of the core libraries of Unikraft, and took the longest to design and implement. Our goal was to enable high-performance networking regardless of the virtualization technology, and enable all other code running on top of `uknetdev` (e.g., lwip) to be sheltered from platform-specific changes.

On KVM, the `uknetdev` API can be configured to use the standard virtio-net protocol and tap devices in the host (vhost-net, the default configuration we used so far), but it can also offload the datapath to vhost-user (a DPDK-based virtio transport running in host userspace) for higher performance – at the cost of polling in the host.

To understand how efficient the `uknetdev` API is, we wrote a simple app that sends as many packets as possible, and measured the achieved throughput on a different machine running a DPDK `testpmd` application. We varied packet sizes and measured throughput, comparing the `uknetdev` API to DPDK running in a Linux VM (DPDK is currently the gold standard for high-performance networking). The results are shown in Figure 19, showing that `uknetdev` with vhost-user offloading achieves similar throughput than DPDK running in a Linux VM.

## 6.3 VFS and Filesystem Specialization (scenario 3)

In this specialization experiment, we aim to obtain high performance out of a web cache application by removing Unikraft's vfs layer (`vfscore`) and hooking the application directly into a purpose-built specialized hash-based filesystem called SHFS, ported from [39].

To benchmark performance, we measure the time it takes to look up a file and open a file descriptor for it. For this purpose, we prepare a small root filesystem with files at the filesystem root. We measure the average time taken to do one open request out of a loop of 1000 open requests, and consider two cases: open() requests where a file exists, and open() requests where it does not. We compare this specialized setup versus running the same application in a Linux VM with an initrd and the files in RAM, and also versus the application running on Unikraft on top of `vfscore` (so no specialization).

The results in Figure 22 show that running the application in Unikraft without specialization already comes with some gains with respect to running it in a Linux VM; however, the big gain comes when running the specialized unikernel:

in this case, we see a 5-7x reduction compared to the non-specialized version, and even higher compared to Linux.

### 6.4 Specializing a key-value store (Scenario 7)

We implemented a UDP-based in-memory key-value store using the `recvmsg/sendmsg` syscalls and then created a lwip-based Unikraft image based on it, as well as Linux binaries.

We measured the request rate each version can sustain, showing the results in Table 4. Unikraft's performance (LWIP) is slightly under that of a Linux guest (Single) and under half of baremetal Linux (Single); nevertheless, performance is low across the board.

| Setup | Mode | Throughput |
|---|---|---|
| Linux baremetal | Single | 769 K/s |
| | Batch | 1.1 M/s |
| Linux guest | Single | 418 K/s |
| | Batch | 627 K/s |
| | DPDK | 6.4 M/s |
| Unikraft guest | LWIP | 319 K/s |
| | `uknetdev` | 6.3 M/s |
| | DPDK | 6.3 M/s |

**Table 4.** Performance of a specialized UDP-based in-memory key-value store on Unikraft vs. Linux.

To improve Linux's performance, we must amortize the cost of system calls (a single packet is sent per syscall in the basic version). To this end we used batched versions of the `msg` syscalls, leading to a roughly 50% improvement in both the baremetal and guest cases. To further improve performance, we ported our app to run on top of DPDK, which requires a major overhaul of the app so that it fits in the DPDK framework—this boosts guest performance to 6.4 million req/s, but at the cost of using two cores in the VM, one exclusively for DPDK.

For Unikraft, we remove the lwip stack and scheduler altogether (via Unikraft's Kconfig menu) and code against the `uknetdev` API, which we use in polling mode. Our specialized unikernel required similar porting effort to the DPDK one. Its performance matches the DPDK performance but it does so using much fewer resources: it only needs one core instead of two for DPDK, it has an image size of 1.4MB compared to about 1GB, and it boots in 80ms instead of a few seconds.

## 7 Discussion

**Do Unikernels trade-off security?** Unikernels traditionally have had serious security issues, but we argue that those have been implementation artifacts rather than fundamental problems. In fact, unikernels are used commercially in security-minded domains such as automotive because their high level of specialization means that they provide a small Trusted Computing Base. Having said that, past unikernel projects have failed to provide standard security features commonly found in standard OSes (e.g., stack and page protection, ASLR, etc.); Unikraft already supports several of these including CFI [3] and Address Sanitisation [70–72], as well as initial support for hardware compartmentalization with Intel MPK [1]. It should therefore be possible to achieve good security while retaining high performance with Unikraft.

**Debugging** One of the common downsides of specialized operating systems is how difficult it is to debug them; unikernels, for instance, are infamous for not having nearly as rich a set of tools for debugging as Linux does.

Unikraft has a number of facilities for helping with this, beyond the fact that its images can of course be run with gdb. First, Unikraft comes with a `ukdebug` micro-library that enables printing of key messages at different (and configurable) levels of criticality. This library can also enable/disable assertions, and print the bottom address of the stack in its messages. Second, Unikraft comes with a trace point system also available through `ukdebug`'s menu options. Third, the `ukdebug` micro-library provides an abstraction to plug in disassemblers; so far, a port of the Zydis disassembler [7] provides such support for x86.

Finally, Unikraft comes with a `linuxu` (Linux user-space) platform target: this allows users, during development, to build their specialized OSes for this target, leveraging Linux's rich set of debugging tools in the process. Once the image runs as expected, users can then choose a different platform, (*e.g.,* Xen) for actual deployment.

**Processes (or lack thereof) in Unikraft.** Unikraft currently does not support processes and their related functions (*e.g.,* `fork()` and `exec()`), although they could be implemented by cloning VMs [81, 82] or emulating them via intra-address space isolation, *e.g.*, Iso-Unik [43] with Intel MPK. Unikraft does have page table support as a micro-library which could be used as the basis to implement `fork` and processes in general. Many modern applications however no longer depend on processes to function [5], and those that do often provide a configurable, thread-based alternative (e.g., nginx).

## 8 Related Work

Over time, a lot of research and systems work has targeted software specialization in various forms; we cover these next.

**Unikernels and Library OSes.** Recent years have seen a large number of library OS or unikernel projects appear. Most are application-specific or programming language-specific projects: runtime.js [65] (JavaScript), includeOS [9] (C++), HaLVM [20] (Haskell), LING or Erlang on Xen [23] (Erlang), MiniPython [49] (MicroPython), ClickOS [51][52] (Click Modular Router/Network Function Virtualization (NFV)), and MiniCache [39][38](content cache). LightVM [48] built tiny unikernels and customized the Xen toolstack to achieve low boot times. In contrast to all of these, Unikraft supports a range of mainstream applications and runtimes efficiently.

MirageOS [46] is an OCaml-specific unikernel focusing on type-safety, so does not support mainstream applications, and its performance, as shown in our evaluation, is sub-par with respect to other unikernel projects and Unikraft. Further, while it provides a rather small kernel (Mini-OS [41]), it is still monolithic, and so not easily customizable.

UniK [29] provides a wrapper around multiple other unikernel projects. Containers and Kata Containers [62] (merged in 2017) [76] provide Linux-based virtual machines (VMs) customized to run containers within them. Drawbridge [59] is a library OS that can run the latest releases of Microsoft applications. Graphene [73] is a library OS targeting efficient execution of single and multi-process applications.

Rump[35] introduces the Anykernel concept, converting parts of the NetBSD kernel to run in a single address space and dealing with issues such as supporting fork and execve. It provides good compatibility for standard applications, but its reliance on a monolithic kernel means there is not much room for specialization. SEUSS [10] uses Rump to build a Function as a Service (FaaS) system. HermiTux [55] is a unikernel providing binary compatibility through syscall re-writing. Unlike Unikraft, HermiTux is not customizable, and its reliance on binary compatibility comes with performance costs, as shown previously. Finally, OSv [37] is an OS focusing on cloud computing that, like HermiTux, comes with binary compatibility support, as well as support for many programming languages and platforms. Its kernel is also monolithic, making it difficult to customize.

**Stack and API Specialization.** A number of works have looked at specializing the software and network stack for various use cases. Sandstorm [50] introduced an extremely (hand-)customized network stack for supporting high performance web and Domain Name System (DNS) servers. The null-Kernel [44], like Unikraft, provides interfaces at different levels of abstraction from the hardware, and allows applications/processes to combine these different interfaces; however, they provide no implementation nor details about application compatibility.

Application-Specific Software Stacks [13] implements compiler extensions to automatically eliminate code even for interpreted languages and shared libraries, and is complementary to Unikraft. Another complementary work is HALO [66], an automated post-link optimization tool that analyzes how memory allocations are made to specialize memory-management routines and thus increase spatial locality.

**Kernel Bypass.** In order to specialize for I/O-bound applications and workloads, a number of works have been proposed that partially or fully bypass the kernel to increase throughput. Demikernel [80] is a new library OS specifically targeting kernel-bypass devices. IX [6] is a novel operating system targeting high throughput and low latency. Arakis [58] is also a new OS where applications have direct access to virtualized I/O devices, while the kernel provides network and disk protection but is not involved in every operation. Parakernel [18], makes bypass a first-class citizen, allowing processes direct access to non-shared devices, and transparently multiplexes access to devices shared by multiple processes. Finally, MDev-NVMe [57] implements pass-through for NVMe SSDs. Unlike these approaches, Unikraft's customizable APIs means that it can achieve high performance without having to bypass the kernel.

**Bridging the Userspace / Kernel Divide.** A number of projects are aimed at running an application and the Linux kernel in a single memory address space in order to reduce domain switch costs, the way unikernels do. User Mode Linux (UML) [15], for instance, allows for running the kernel as a user space process. LibOS [26] runs the kernel network stack as a shared library in user-space. Linux Kernel Library (LKL) [60] transforms the kernel into a library that can be run as a virtual machine. Unikernel Linux (UKL) [63] takes a user-space application and statically compiles it into the Linux kernel, using a shim layer to redirect `glibc` syscalls to kernel functions (but adds 45MB of `glibc` code to every image). These approaches avoid user-kernel space switches, but do not specialize the actual software stack. More recent work called Lupine [40] uses the Linux kernel to load an application binary in kernel mode, thus eliminating the system call overhead; it also makes use of kernel configuration options to discard unnecessary features, thus reducing the image size. We have shown, however, that it underperforms when compared to Unikraft.

## 9 Conclusions

We have introduced Unikraft, a novel micro-library OS targeting high performance through full and easy specialization of unikernel images. In addition to yielding benefits to unmodified applications (e.g., faster boot times, lower memory consumption, etc.), slight modifications to applications to comply with Unikraft's APIs result in even higher performance.

As future work, we are continuing the effort to provide better syscall compatibility in order to transparently support even more mainstream applications. We also aim to leverage Unikraft's modularity for security purposes, coding micro-libraries in memory-safe or even statically-verifiable languages and using compartmentalization techniques to maintain safety properties as the image is linked together.

## Acknowledgments

We would like to thank the anonymous reviewers and our shepherd, Manuel Costa, for their comments and insights. A very special thank you goes to the Unikraft OSS community for their substantial past and ongoing contributions. This work was partly funded by EU H2020 grant agreements 825377 (UNICORE) and 871793 (ACCORDION). Costin Raiciu was partly supported by gift funding from VMWare.

# Artifact Appendix

We have spent a considerable effort trying to ensure that all of the experiments in this paper are reproducible. This appendix provides thorough information on how to run them; the Artifact Evaluation (AE) repo can be found at [22]. In addition, we prove tables summarizing all experiments with a rough estimate of how much time it takes for each to run.

## 9.1 Hardware Requirements

Before you can run these experiments, you will need to prepare 3 physical host environments: physical hosts as opposed to virtual machines are recommended as they provide better performance. In the paper, we used three different setups:

1. A Linux host (Debian Buster) with KVM enabled and Linux kernel 4.19. This host is used for most experiments. We use the 4.19 kernel because HermiTux will not run with newer versions, as noted in [27].
2. A Linux host (Debian Buster) with Linux kernel 4.19 that has an 10gbit/s Ethernet cable connected to the first host. We use it for the DPDK network experiments Figure 19 and Table 4 and experiments where we need to specifically setup the CPU frequency. See Section 9.2 for further details.
3. Xen host (Debian Buster) used for Xen 9pfs experiments Figure 20.

A single server can be used for almost all experiments, though it would require different Linux kernel parameters, or the Xen hypervisor and rebooting to switch from one set up to another. The exception is the DPDK experiment, which requires two servers connected to each other via a 10Gb link.

As a reminder, all of our results were run on inexpensive (roughly €800) Shuttle SH370R6 [68] computer with an Intel i7 9700K 3.6 GHz (4.9 Ghz with Turbo Boost, 8 cores) and 32GB of RAM. For the DPDK experiment we use two of these connected via a direct cable and a pair of Intel X520-T2 [33] cards with the 82599EB chipset.

## 9.2 Software Requirements

All experiments were run on a physical host with Debian Buster and Linux 4.19 installed. All install and preparation scripts in the AE repository target this distribution and kernel version.

For all set ups, we disabled Hyper-Threading (`noht`), isolated 4 CPU cores (e.g. `isocpus=2-6`), switched off the IOMMU (`intel_iommu=off`), and disabled IPv6 (`ipv6.disable=1`). This can be done by setting kernel boot parameters with your bootloader, for instance with Grub (`/etc/default/grub`):

```
GRUB_CMDLINE_LINUX_DEFAULT="isolcpus=2-4    \
                            noht            \
                            intel_iommu=off \
                            ipv6.disable=1"
```

or with syslinux/pxelinux:

| Figure | Description | Time |
|---|---|---|
| Fig 1 | Linux kernel dependency graph | 0h 50m |
| Fig 2 | NGINX Unikraft dependency graph | 0h 5m |
| Fig 3 | "Hello World" Unikraft dependency graph | 0h 1m |
| Fig 5 | Syscalls required by a set of 30 popular server applications versus syscalls currently supported by Unikraft | 0h 45m |
| Fig 7 | Syscall support for top 30 server apps. All apps are close to being supported, and several already work even if some syscalls are stubbed (SQLite, NGINX) | 0h 45m |
| Fig 8 | Image sizes of Unikraft applications. We include permutations with and without LTO and DCE | 0h 1m |
| Fig 9 | Image sizes for representative applications with Unikraft and other OSes, stripped, without LTO and DCE | 0h 5m |
| Fig 10 | Boot time for Unikraft images with different virtual machine monitor | 0h 9m |
| Fig 11 | Minimum memory needed to run different applications using different OSes, including Unikraft | 0h 50m |
| Fig 12 | Redis performance tested with the redis-benchmark, (30 connections, 100k requests, pipelining level of 16) | 0h 9m |
| Fig 13 | NGINX (and Mirage HTTP-reply) performance tested with wrk (1 minute, 14 threads, 30 conns, static 612B HTML page) | 0h 50m |
| Fig 14 | Unikraft Boot time for NGINX with different memory allocators | 0h 8m |
| Fig 15 | NGINX throughput with different memory allocators | 0h 30m |
| Fig 16 | Execution speedup in SQLite Unikraft, relative to mimalloc | 0h 21m |
| Fig 17 | Time for 60k SQLite insertions with native Linux,newlib and musl on Unikraft (marked as native) and SQLite ported automatically to Unikraft (musl external) | 0h 6m |
| Fig 18 | Throughput for Redis Unikraft, with varying memory allocators and request type | 0h 5m |
| Fig 19 | TX throughput comparison of Unikraft versus DPDK in a Linux VM | 0h 30m |
| Fig 20 | 9pfs latency for read and write operations, compared to Linux | 2h 0m |
| Fig 21 | Unikraft boot times with static and dynamic initialization of page tables | 0h 3m |
| Fig 22 | Filesystem specialization and removal of the vfs layer yields important performance gains for a web cache | 0h 5m |

| Table | Description | Time |
|---|---|---|
| Tab 1 | Cost of binary compatibility/syscalls with and without security mitigations | 0h 25m |
| Tab 2 | Results from automated porting based on externally-built archives when linked against Unikraft using musl and newlib. We show whether the port succeeded with the glibc compatibility layer ("compat layer") and without it ("std"). | 0h 25m |
| Tab 4 | Performance of a specialized UDP-based in-memory key-value store on Unikraft vs. Linux | 0h 25m |

| Text | Description | Time |
|---|---|---|
| Text 1 | Unikernel boot time baseline | 0h 21m |
| Text 2 | Measures 9pfs boot time overhead | 0h 5m |

```
LABEL item_kernel0
  MENU LABEL Linux
  MENU DEFAULT
  KERNEL vmlinuz-4.19.0
 APPEND isolcpus=2-6 noht intel_iommu=off ipv6.disable=1
```

On Xen we use the following parameters (please adjust the amount of pinned memory for Dom0 according to your available RAM, we gave the half of 32GB RAM to Dom0; We also pinned 4 CPU cores to Dom0): Grub (`/etc/default/grub`):

```
GRUB_CMDLINE_LINUX_XEN_REPLACE_DEFAULT=""
GRUB_CMDLINE_LINUX_XEN_REPLACE="earlyprintk=xen \
                                console=hvc0 \
                                ipv6.disable=1"
GRUB_CMDLINE_XEN=""
GRUB_CMDLINE_XEN_DEFAULT="smt=0 dom0_max_vcpus=4 \
                          dom0_vcpus_pin cpufreq=xen \
                          dom0_mem=15360M,max:16384M \
                          gnttab_max_frames=256"
```

Please note that the following experiments require additional kernel parameters e.g., to enable specific CPU frequency scaling governors: tables 1 and 4 and figs. 19 and 22.

### 9.3 Getting Started

1. Before running any of these experiments, prepare your host with the recommendations detailed above.
2. Many of the experiments use Docker as an intermediate tool for creating build and test environments (along with testing Docker itself). Please install Docker [16] on your system.
3. Once Docker is installed, clone our AE repository:

   ```
   git clone \
   https://github.com/unikraft/eurosys21-artifacts.git
   ```
4. All experiments should be `prepared` first, which installs necessary tools and downloads additional resources, before they can run. This can be done by calling `run.sh fig_XX prepare` (more details below) for a single experiment or `run.sh prepare` for all experiments. (Note: The preparation step for all experiments usually exceeds several hours.)
5. Once prepared, simply call the relevant experiment you wish to re-create using the `run.sh` script.

We have wrapped all the individual experiments with the `run.sh` tool. This script will install the necessary dependencies for all experiments (excluding Docker) for Debian Buster. Each experiment, and more specifically its sub-directory in `experiments/`, is populated with a `README.md` which includes more details about the individual experiment.

### 9.4 Notes

- All experiments should be run as the `root` user on the host as it will require modifications to the host and running commands with elevated privileges, e.g. creating and destroying VMs, setting limits in `/proc`, etc.
- We use intermediate Docker containers for building images and accessing pre-built binaries for many of the experiments. In addition to this, this repository clones the Linux kernel to make changes for testing. As a result, expected disk storage utilized to conduct all experiments is 50GB.
- The preparation step for all experiments usually exceeds several hours.
- Experiments cannot be run in parallel due to overlapping CPU core affinities, which will affect measurements.
- Each experiment has its own sub-directory and a `Makefile` script within it. We further provide a main `run.sh` script that wraps all experiments.
- Some experiments (e.g., Figure 22) produce some error messages but still finish and correctly produce the plot; if this is the case, this is documented in an experiment's sub-directory, in its own `README.md` file.
- All plots are saved into the global `/plots` directory when run via `run.sh`. When using the individual experiment's `Makefile`, it is saved to the experiment's folder.

### 9.5 Beyond the Paper

The AE repository only contains the performance evaluation of Unikraft. In addition to this appendix and the repo, the Unikraft project provides extensive documentation [75] on how to use Unikraft in real-world environments. In addition, interested researchers are welcome to join the community via the Xen project mailing list [79] and GitHub [21].

## References

[1] Intel® 64 and IA-32 Architectures Software Developer's Manual. Volume 3A, Section 4.6.2.

[2] Newlib: a C library intended for use on embedded systems. https://sourceware.org/newlib/. Online; accessed Jan, 25 2021.

[3] Martín Abadi, Mihai Budiu, Úlfar Erlingsson, and Jay Ligatti. Control-flow integrity. In *Proceedings of the 12th ACM Conference on Computer and Communications Security*, CCS '05, pages 340–353, New York, NY, USA, 2005. Association for Computing Machinery.

[4] Alexandru Agache, Marc Brooker, Alexandra Iordache, Anthony Liguori, Rolf Neugebauer, Phil Piwonka, and Diana-Maria Popa. Firecracker: Lightweight virtualization for serverless applications. In *Proceedings of the 17th USENIX Symposium on Networked Systems Design and Implementation*, NSDI'20), pages 419–434, 2020.

[5] Andrew Baumann, Jonathan Appavoo, Orran Krieger, and Timothy Roscoe. A fork() in the road. In *Proceedings of the Workshop on Hot Topics in Operating Systems*, HotOS'19, page 14–22, New York, NY, USA, 2019. Association for Computing Machinery.

[6] Adam Belay, George Prekas, Ana Klimovic, Samuel Grossman, Christos Kozyrakis, and Edouard Bugnion. IX: A protected dataplane operating system for high throughput and low latency. In *Proceedings of the 11th USENIX Symposium on Operating Systems Design and Implementation*, OSDI'14, pages 49–65, Broomfield, CO, 2014. USENIX Association.

[7] Florian Bernd and Joel Höner. Zydis: Fast and lightweight x86/x86-64 disassembler library. https://zydis.re/. Online; accessed Jan, 25 2021.

[8] Ivan T. Bowman, Richard C. Holt, and Neil V. Brewster. Linux as a case study: Its extracted software architecture. In *Proceedings of the 21st International Conference on Software Engineering*, ICSE '99, page 555–563, New York, NY, USA, 1999. Association for Computing Machinery.

[9] Alfred Bratterud, Alf-Andre Walla, Harek Haugerud, Paal E. Engelstad, and Kyrre Begnum. IncludeOS: A minimal, resource efficient unikernel for cloud services. In *Proceedings of the 7th IEEE International Conference on Cloud Computing Technology and Science*, CloudCom'15. IEEE, November 2015.

[10] James Cadden, Thomas Unger, Yara Awad, Han Dong, Orran Krieger, and Jonathan Appavoo. SEUSS: Rapid serverless deployment using environment snapshots. *CoRR*, abs/1910.01558, 2019.

[11] Jonathan Corbet. The rapid growth of io_uring. https://lwn.net/Articles/810414/. Online; accessed Jan, 25 2021.

[12] Thurston H.Y. Dang, Petros Maniatis, and David Wagner. Oscar: A practical page-permissions-based scheme for thwarting dangling pointers. In *Proceedings of the 26th USENIX Security Symposium*, USENIX Security'17, pages 815–832, Vancouver, BC, 2017. USENIX Association.

[13] Nicolai Davidsson, Andre Pawlowski, and Thorsten Holz. Towards automated application-specific software stacks. In *Proceedings of the 24th European Symposium on Research in Computer Security*, pages 88–109, 2019.

[14] Debian. Debian Popularity Contest. https://popcon.debian.org/. Online; accessed Jan, 25 2021.

[15] Jeff Dike. A user-mode port of the linux kernel. In *Proceedings of the 4th Annual Linux Showcase and Conference (Volume 4)*, ALS'00, pages 7–7, Berkeley, CA, USA, 2000. USENIX Association.

[16] Docker Docs. Get Docker. https://docs.docker.com/get-docker/. Online; accessed March, 26 2021.

[17] Adam Dunkels. Design and implementation of the lwip stack. 2001.

[18] Pekka Enberg, Ashwin Rao, and Sasu Tarkoma. I/O Is Faster Than the CPU: Let's Partition Resources and Eliminate (Most) OS Abstractions. In *Proceedings of the Workshop on Hot Topics in Operating Systems*, HotOS'19, pages 81–87, New York, NY, USA, 2019. ACM.

[19] Dario Faggioli. Virtual-machine scheduling and scheduling in virtual machines. https://lwn.net/Articles/793375/, July 2019. Online; accessed Jan, 25 2021.

[20] Galois Inc. The haskell lightweight virtual machine (halvm). https://github.com/GaloisInc/HaLVM, 2008. Online; accessed Jan, 25 2021.

[21] GitHub. A Unikernel SDK. Extreme Specialization for Security and Performance. https://github.com/unikraft. Online; accessed March, 26 2021.

[22] GitHub. Artifacts, including experiments and graphs, for the paper: "Unikraft: Fast, Specialized Unikernels the Easy Way" (EuroSys'21). https://github.com/unikraft/eurosys21-artifacts. Online; accessed March, 26 2021.

[23] GitHub. Erlang on Xen. https://github.com/cloudozer/ling. Online; accessed Jan, 25 2021.

[24] GitHub. Firecracker GitHub issue #1034: Slower networking of OSv on firecracker vs QEMU/KVM. https://github.com/firecracker-microvm/firecracker/issues/1034. Online; accessed Jan, 25 2021.

[25] GitHub. Hermitux GitHub issue #2: It does not work on qemu. https://github.com/ssrg-vt/hermitux/issues/2. Online; accessed Jan, 25 2021.

[26] Github. linux-libos-tools. https://github.com/libos-nuse/linux-libos-tools. Online; accessed Jan, 25 2021.

[27] GitHub. Performance issue with Redis on recent Linux kernels. https://github.com/ssrg-vt/hermitux/issues/12. Online; accessed March, 26 2021.

[28] Github. Xen Minimal OS - Memory management related functions. https://github.com/sysml/mini-os/blob/master/mm.c. Online; accessed Jan, 25 2021.

[29] Github.com. The Unikernel and MicroVM Compilation and Deployment Platform. https://github.com/solo-io/unik. Online; accessed Jan, 25 2021.

[30] Google. Cloud TPU – Train and run machine learning models faster than ever before. https://cloud.google.com/tpu. Online; accessed Jan, 25 2021.

[31] Google. Protocol Buffers – Google's data interchange format. https://github.com/protocolbuffers/protobuf.

[32] Habana. 100% AI. https://habana.ai/. Online; accessed Jan, 25 2021.

[33] Intel. Ethernet-Converged-Network-Adapter X520-T2. https://ark.intel.com/content/www/de/de/ark/products/69655/intel-ethernet-converged-network-adapter-x520-t2.html. Online; accessed Mar, 26 2021.

[34] Intel. Intel® Movidius™ Vision Processing Units (VPUs). https://www.intel.com/content/www/us/en/products/processors/movidius-vpu.html. Online; accessed Jan, 25 2021.

[35] Antti Kantee. Kernel file systems as userspace programs. September 2007.

[36] Antti Kantee. *Flexible Operating System Internals: The Design and Implementation of the Anykernel and Rump Kernels*. PhD thesis, Aalto University, 2012.

[37] Avi Kivity, Dor Laor, Glauber Costa, Pekka Enberg, Nadav Har'El, Don Marti, and Vlad Zolotarov. OSv–Optimizing the Operating System for Virtual Machines. In *Proceedings of the 2014 USENIX Annual Technical Conference*, USENIX ATC'14, pages 61–72, Philadelphia, PA, June 2014. USENIX Association.

[38] Simon Kuenzer, Anton Ivanov, Filipe Manco, Jose Mendes, Yuri Volchkov, Florian Schmidt, Kenichi Yasukata, Michio Honda, and Felipe Huici. Unikernels everywhere: The case for elastic cdns. In *Proceedings of the 13th ACM SIGPLAN/SIGOPS International Conference on Virtual Execution Environments*, VEE '17, pages 15–29, New York, NY, USA, 2017. ACM.

[39] Simon Kuenzer, Joao Martins, Mohamed Ahmed, and Felipe Huici. Towards minimalistic, virtualized content caches with minicache. In *Proceedings of the 2013 ACM Workshop on Hot Topics in Middleboxes and Network Function Virtualization*, HotMiddlebox'13, pages 13–18. ACM, 2013.

[40] Hsuan-Chi Kuo, Dan Williams, Ricardo Koller, and Sibin Mohan. A linux in unikernel clothing. In *Proceedings of the Fifteenth European Conference on Computer Systems*, EuroSys '20, New York, NY, USA, 2020. Association for Computing Machinery.

[41] Lars Kurth and Russell Pavlicek. Xen Project Wiki Mini-OS. https://wiki.xenproject.org/wiki/Mini-OS, 2018. Online; accessed Jan, 25 2021.

[42] Daan Leijen, Benjamin Zorn, and Leonardo de Moura. Mimalloc: Free list sharding in action. In *Asian Symposium on Programming Languages and Systems*, pages 244–265. Springer, 2019.

[43] Guanyu Li, Dong Du, and Yubin Xia. Iso-unik: lightweight multi-process unikernel through memory protection keys. *Cybersecur.*, 3(1):11, 2020.

[44] James Litton, Deepak Garg, Peter Druschel, and Bobby Bhattacharjee. Composing abstractions using the null-kernel. In *Proceedings of the Workshop on Hot Topics in Operating Systems*, HotOS'19, pages 1–6, New York, NY, USA, 2019. ACM.

[45] Anil Madhavapeddy, Thomas Leonard, Magnus Skjegstad, Thomas Gazagnaire, David Sheets, Dave Scott, Richard Mortier, Amir Chaudhry, Balraj Singh, Jon Ludlam, Jon Crowcroft, and Ian Leslie. Jitsu: Just-In-Time Summoning of Unikernels. In *12th USENIX Symposium on Networked Systems Design and Implementation*, NSDI '15, pages 559–573, Oakland, CA, 2015. USENIX Association.

[46] Anil Madhavapeddy, Richard Mortier, Charalampos Rotsos, David Scott, Balraj Singh, Thomas Gazagnaire, Steven Smith, Steven Hand, and Jon Crowcroft. Unikernels: Library operating systems for the cloud. In *Proceedings of the 18th International Conference on Architectural Support for Programming Languages and Operating Systems (ASPLOS'13)*. ACM, 2013.

[47] Anil Madhavapeddy and David J. Scott. Unikernels: Rise of the Virtual Library Operating System. *Queue*, 11(11):30:30–30:44, December 2013.

[48] Filipe Manco, Costin Lupu, Florian Schmidt, Jose Mendes, Simon Kuenzer, Sumit Sati, Kenichi Yasukata, Costin Raiciu, and Felipe Huici. My vm is lighter (and safer) than your container. In *Proceedings of the 26th Symposium on Operating Systems Principles*, SOSP '17, pages 218–233, New York, NY, USA, 2017. ACM.

[49] Filipe Manco, Costin Lupu, Florian Schmidt, Jose Mendes, Simon Kuenzer, Sumit Sati, Kenichi Yasukata, Costin Raiciu, and Felipe Huici. My VM is lighter (and safer) than your container. In *Proceedings of the 26th ACM Symposium on Operating Systems Principles*, SOSP'17. ACM, 2017.

[50] Ilias Marinos, Robert N.M. Watson, and Mark Handley. Network Stack Specialization for Performance. In *Proceedings of the 2014 ACM Conference on Computer Communication*, SIGCOMM '14, pages 175–186, New York, NY, USA, 2014. ACM.

[51] Joao Martins, Mohamed Ahmed, Costin Raiciu, and Felipe Huici. Enabling fast, dynamic network processing with clickos. In *Proceedings of the 2nd ACM SIGCOMM Workshop on Hot Topics in Software Defined Networking*, HotSDN"13. ACM, 2013.

[52] Joao Martins, Mohamed Ahmed, Costin Raiciu, Vladimir Olteanu, Michio Honda, Roberto Bifulco, and Felipe Huici. ClickOS and the art of network function virtualization. In *Proceedings of the 11th USENIX Conference on Networked Systems Design and Implementation*, NSDI'14, pages 459–473. USENIX, 2014.

[53] Miguel Masmano, Ismael Ripoll, Alfons Crespo, and Jorge Real. TLSF: A new dynamic memory allocator for real-time systems. In *Proceedings of the 16th Euromicro Conference on Real-Time Systems*, ECRTS'04, pages 79–88. IEEE, 2004.

[54] Pierre Olivier, Daniel Chiba, Stefan Lankes, Changwoo Min, and Binoy Ravindran. A binary-compatible unikernel. In *Proceedings of the 15th ACM SIGPLAN/SIGOPS International Conference on Virtual Execution Environments*, VEE 2019, pages 59–73, New York, NY, USA, 2019. ACM.

[55] Pierre Olivier, Daniel Chiba, Stefan Lankes, Changwoo Min, and Binoy Ravindran. A binary-compatible unikernel. In *Proceedings of the 15th ACM SIGPLAN/SIGOPS International Conference on Virtual Execution Environments (VEE)*, VEE'19, pages 59–73. ACM, 2019.

[56] Openwall. Implement glibc chk interfaces for ABI compatibility. https://www.openwall.com/lists/musl/2015/06/17/1. Online; accessed Jan, 25 2021.

[57] Bo Peng, Haozhong Zhang, Jianguo Yao, Yaozu Dong, Yu Xu, and Haibing Guan. Mdev-nvme: A nvme storage virtualization solution with mediated pass-through. In *Proceedings of the 2018 USENIX Conference on Usenix Annual Technical Conference*, USENIX ATC'18, page 665–676, USA, 2018. USENIX Association.

[58] Simon Peter, Jialin Li, Irene Zhang, Dan R. K. Ports, Doug Woos, Arvind Krishnamurthy, Thomas Anderson, and Timothy Roscoe. Arrakis: The operating system is the control plane. In *11th USENIX Symposium on Operating Systems Design and Implementation (OSDI 14)*, pages 1–16, Broomfield, CO, October 2014. USENIX Association.

[59] Donald E. Porter, Silas Boyd-Wickizer, Jon Howell, Reuben Olinsky, and Galen C. Hunt. Rethinking the library os from the top down. In *Proceedings of the Sixteenth International Conference on Architectural Support for Programming Languages and Operating Systems*, ASPLOS'16, page 291–304, New York, NY, USA, 2011. Association for Computing Machinery.

[60] Octavian Purdila, Lucian Grijincu, and Nicolae Tapus. Lkl: The linux kernel library. Proceedings of the Roedunet International Conference, pages 328 – 333, 07 2010.

[61] Anh Quach, Rukayat Erinfolami, David Demicco, and Aravind Prakash. A multi-os cross-layer study of bloating in user programs, kernel and managed execution environments. In *Proceedings of the 2017 Workshop on Forming an Ecosystem Around Software Transformation*, FEAST '17, page 65–70, New York, NY, USA, 2017. Association for Computing Machinery.

[62] Alessandro Randazzo and Ilenia Tinnirello. Kata containers: An emerging architecture for enabling mec services in fast and secure way. In *Proceedings of the 6th International Conference on Internet of Things: Systems, Management and Security*, IOTSMS'19, pages 209–214. IEEE, 2019.

[63] Ali Raza, Parul Sohal, James Cadden, Jonathan Appavoo, Ulrich Drepper, Richard Jones, Orran Krieger, Renato Mancuso, and Larry Woodman. Unikernels: The next stage of linux's dominance. In *Proceedings of the Workshop on Hot Topics in Operating Systems*, HotOS '19, pages 7–13, New York, NY, USA, 2019. ACM.

[64] Luigi Rizzo. netmap: A novel framework for fast packet I/O. In *Proceedings of the 21st USENIX Annual Technical Conference*, USENIX ATC'12, pages 101–112. USENIX, 2012.

[65] runtimejs.org. JavaScript Library Operating System for the Cloud. http://runtimejs.org/. Online; accessed Jan, 25 2021.

[66] Joe Savage and Timothy M. Jones. Halo: Post-link heap-layout optimisation. In *Proceedings of the 18th ACM/IEEE International Symposium on Code Generation and Optimization*, CGO'20, page 94–106, New York, NY, USA, 2020. Association for Computing Machinery.

[67] Karsten Schmidt. malloc, free replacement for unmanaged, linear memory situations. https://github.com/thi-ng/tinyalloc. Online; accessed Jan, 25 2021.

[68] Shuttle. SH370R6 XCP Cube. http://global.shuttle.com/products/productsDetail?productId=2265. Online; accessed Mar, 26 2021.

[69] Mincheol Sung, Pierre Olivier, Stefan Lankes, and Binoy Ravindran. Intra-unikernel isolation with intel memory protection keys. In *Proceedings of the 16th ACM SIGPLAN/SIGOPS International Conference on Virtual Execution Environments*, VEE '20, page 143–156, New York, NY, USA, 2020. Association for Computing Machinery.

[70] The Linux Kernel Development Community. The kernel address sanitizer (KASAN). https://www.kernel.org/doc/html/v5.10/dev-tools/kasan.html. Online; accessed Jan, 25 2021.

[71] The Linux Kernel Development Community. The kernel concurrency sanitizer (KCSAN). https://www.kernel.org/doc/html/v5.10/dev-tools/kcsan.html. Online; accessed Jan, 25 2021.

[72] The Linux Kernel Development Community. The undefined behavior sanitizer (UBSAN). https://www.kernel.org/doc/html/v5.10/dev-tools/ubsan.html. Online; accessed Jan, 25 2021.

[73] Chia-Che Tsai, Kumar Saurabh Arora, Nehal Bandi, Bhushan Jain, William Jannen, Jitin John, Harry A. Kalodner, Vrushali Kulkarni, Daniela Oliveira, and Donald E. Porter. Cooperation and security isolation of library oses for multi-process applications. In *Proceedings of the 9th European Conference on Computer Systems*, EuroSys'14, pages 9:1–9:14, New York, NY, USA, 2014. ACM.

[74] Chia-Che Tsai, Bhushan Jain, Nafees Ahmed Abdul, and Donald E. Porter. A study of modern linux api usage and compatibility: What to support when you're supporting. In *Proceedings of the Eleventh European Conference on Computer Systems*, EuroSys '16, New York, NY, USA, 2016. Association for Computing Machinery.

[75] unikraft.org. Unikraft's Documentation. http://docs.unikraft.org/. Online; accessed March, 26 2021.

[76] Arjan Van de Ven. An introduction to Clear Containers. https://lwn.net/Articles/644675/. Online; accessed Jan, 25 2021.

[77] Eric Van Hensbergen and Ron Minnich. Grave robbers from outer space using 9p2000 under linux. In *Proceedings of the USENIX Annual Technical Conference*, ATC'05, page 45, USA, 2005. USENIX Association.

[78] Dan Williams and Ricardo Koller. Unikernel Monitors: Extending Minimalism Outside of the Box. In *8th USENIX Workshop on Hot Topics in Cloud Computing*, HotCloud '16, Denver, CO, 2016. USENIX Association.

[79] Xen Project. Minios-devel – Mini-os development list. https://lists.xenproject.org/cgi-bin/mailman/listinfo/minios-devel. Online; accessed March 26, 2021.

[80] Irene Zhang, Jing Liu, Amanda Austin, Michael Lowell Roberts, and Anirudh Badam. I'm not dead yet!: The role of the operating system in a kernel-bypass era. In *Proceedings of the Workshop on Hot Topics in Operating Systems*, HotOS'19, pages 73–80, New York, NY, USA, 2019. ACM.

[81] Yiming Zhang, Jon Crowcroft, Dongsheng Li, Chengfen Zhang, Huiba Li, Yaozheng Wang, Kai Yu, Yongqiang Xiong, and Guihai Chen. KylinX: A Dynamic Library Operating System for Simplified and Efficient Cloud Virtualization. In *2018 USENIX Annual Technical Conference*, USENIX ATC'18, pages 173–186. USENIX Association, July 2018.

[82] Yiming Zhang, Chengfei Zhang, Yaozheng Wang, Kai Yu, Guangtao Xue, and Jon Crowcroft. Kylinx: Simplified virtualization architecture for specialized virtual appliances with strong isolation. *ACM Trans. Comput. Syst.*, 37(1–4), February 2021.